\documentclass[a4paper,12pt]{article}
\usepackage[english]{babel}
\usepackage[colorlinks,breaklinks]{hyperref}
\hypersetup{
    colorlinks,%
    citecolor=black,%
    filecolor=black,%
    linkcolor=black,%
    urlcolor=black
}

\usepackage[table, x11names]{xcolor}
\usepackage{url}
\usepackage{layaureo}
\usepackage{fullpage}

\usepackage{graphicx}
\usepackage{color}
\usepackage[nomarkers,nolists]{endfloat}
\usepackage{multirow}

\usepackage{amsthm}
\usepackage{amsmath}
\usepackage{amsfonts}
\usepackage{amssymb}
\usepackage{graphicx}
\usepackage{setspace}
\usepackage{natbib}
\usepackage{booktabs}
\usepackage{array}
\newcolumntype{K}[1]{>{\centering\let\newline\\\arraybackslash\hspace{0pt}}m{#1}}

\usepackage{lscape}
\usepackage{ctable}

\usepackage{array}
\newcolumntype{C}[1]{>{\centering\let\newline\\\arraybackslash\hspace{0pt}}m{#1}}

\usepackage{xcolor,colortbl}

\numberwithin{equation}{section}

\title{}
\author{
\\
\\
}

\begin{document}

\begin{center}
{\noindent \textbf{\Large CoVaR with volatility clustering, heavy tails\\ and non-linear dependence}}

\vspace{14pt}

{\noindent Michele Leonardo Bianchi,\textsuperscript{a,}\footnote{ This publication should not be reported as representing the views of the Bank of Italy. The views expressed are those of the authors and do not necessarily reflect those of the Bank of Italy.} Giovanni De Luca\textsuperscript{b}, and Giorgia Rivieccio\textsuperscript{b}}

\vspace{14pt}

{\noindent \small\textsuperscript{a}\it Regulation and Macroprudential Analysis Directorate,
Bank of Italy, Rome, Italy}\\
{\noindent \small\textsuperscript{b}\it Department of Management and Quantitative Studies,
Univerisity of Naples Parthenope, Naples, Italy}\\

\vspace{20pt}

This version: \today

\end{center}

\vspace{14pt}

\noindent {\bf Abstract.}  In this paper we estimate the conditional value-at-risk by fitting different multivariate parametric models capturing some stylized facts about multivariate financial time series of equity returns: heavy tails, negative skew, asymmetric dependence, and volatility clustering. While the volatility clustering effect is got by AR-GARCH dynamics of the GJR type, the other stylized facts are captured through non-Gaussian multivariate models and copula functions. The CoVaR$^{\leq}$ is computed on the basis on the multivariate normal model, the multivariate normal tempered stable (MNTS) model, the multivariate generalized hyperbolic model (MGH) and four possible copula functions. These risk measure estimates are compared to the CoVaR$^{=}$ based on the multivariate normal GARCH model. The comparison is conducted by backtesting the competitor models over the time span from January 2007 to March 2020. In the empirical study we consider a sample of listed banks of the euro area belonging to the main or to the additional global systemically important banks (GSIBs) assessment sample.

\vskip 0.6cm


\section{Introduction}

The contribution of financial institutions to systemic risks in financial markets is a largely debated topic in literature.
Relevant articles include \cite{billio}, \cite{adrian2016covar}, \cite{girardi2013covar}, \citet{Lin}.

In \citet*{adrian2016covar} the distress of a financial institution is defined as the event $(y^j = \mbox{VaR}_{\alpha}^j)$, where $y^j$ is the random variable representing the log-returns of the financial institution $j$ and $\mbox{VaR}_{\alpha}^j$ the corresponding value-at-risk (VaR) at tail level $\alpha$.  Here we consider the conditional value-at-risk (CoVaR$^{\leq}$) measure, that is the conditional value-at-risk where the conditioning event is the distress of a financial institution represented through the inequality $y^j \leq \mbox{VaR}_{\alpha}^j$. This allows us to have a robust systemic risk measure which can be backtested without a great effort (see \cite{girardi2013covar} and \cite{banulescu2019backtesting}). Conversely the original CoVaR of \cite{adrian2016covar} is simple to estimate, as shown in \cite{bs2019covar}, but not so simple to backtest.

Assuming that the CoVaR$^{\leq}$ is a proper systemic risk measure, we explore to which extent the model assumptions on the univariate financial institution log-returns and on the dependence structure affect the estimates of this risk measure. Additional, we conduct a backtesting analysis to obtain a robust model comparison.

It is widely known that the amplitude of daily returns varies over time and that if the volatility is high, it tends to remain high, and if it is low, it tends to remain low. This means that volatility moves in clusters and for this reason it is necessary to capture such observed behaviour (\cite{rachev2011financial}). Additionally the CoVaR estimation is by definition a multi-dimensional problem. The multivariate normal model is usually applied in practical applications to finance, mainly because both the theoretical and practical complexity of a model increases if one moves from a normal to a non-normal framework. 
However, the multivariate normal distribution has two main drawbacks: (1) its margins are normally distributed, therefore it does not capture empirically observed skewness and kurtosis; (2) its dependence structure is symmetric, it does not capture asymmetry of dependence during extreme market movements and the dependence of tail events. For the reasons above, in this work we implement multivariate non-normal models with volatility clustering for the CoVaR estimation.

More in details, we assume that the univariate time series have AR-GARCH dynamics with Glosten-Jagannathan-Runkle (GJR) volatility (see \cite{glosten1993relation}) and then we analyze different dependence structures. 
Among possible multivariate parametric models applied to finance (see \cite{bhmrt2020multivariate}), we select the multivariate normal tempered stable (MNTS) and the multivariate generalized hyperbolic (MGH) model, and four copula functions, normal, $t$ and BB1 and BB7, as described in \cite{jaworski2017conditional}, \cite{deluca2018covar} and \cite{deluca2019var}. Both non-normal multivariate distributions and copula functions are widely known in the financial literature. \cite{btf2015riding} and \cite{bianchi2017forward} analyzed both MNTS e MGH models applied to risk assessment and portfolio optimizations (see also \cite{fallahgoul2019modelling} and \cite{handbook2019bstff}). \cite{kurosaki2013sytematic} developed a model based on the MNTS distribution to estimate the CoVaR, and subsequently \cite{kurosaki2013mean} and \cite{biglova2014portfolio} studied a mean-CoAVaR strategy applied to portfolio optimization to mitigate the potential loss arising from systemic risk.

The remainder of the paper is organized as follows. In Section \ref{sec:Methodology} we describe the methodology implemented in this work: we define both the univariate model and the dependence structure, we show the necessary formulas to compute the CoVaR$^{\leq}$. After having described in Section \ref{sec:Data} the market data considered in this study, the main empirical results are discussed in Section \ref{sec:Results}. In Section \ref{sec:Backtest} we compare the different distributional assumptions through a backtesting exercise. In Section \ref{sec:GSIBs} we compare the $\Delta \mbox{CoVaR}$ with the score defined by the Financial Stability Board (FSB) for the global systemically important banks (GSIBs) bucket allocation and we introduce a score adjusted for the information coming from the stock markets.. Section \ref{sec:Conclusions} concludes.

\section{Methodology}\label{sec:Methodology}

For each institution $j$, the random variable $y_t^j$ represents the log-returns of the market value of equity. Superscript {\it sys} denotes the entire financial system, i.e. the capitalization-weighted portfolio of all financial institutions in the selected sample or an index representative of the stock market and frequently used by financial professionals (i.e. the S\&P 500 index or the Euro Stoxx 50 index).

At time $t$, given the VaR of the financial institution $j$, with tail level $\alpha$ ($\mbox{VaR}_{\alpha,t}^j$), for a given tail level $\beta$, the ${\mbox{CoVaR}^{\leq}}_{\beta,\alpha,t}^{j}$ of the financial system conditional on financial institution $j$ being in distress (i.e. market returns of bank $j$ are less or equal to its $\mbox{VaR}_{\alpha}$) is equal to 
\begin{equation}\label{eq:CoVaR}
P\left(y_t^{sys}\leq {\mbox{CoVaR}^{\leq}}_{\beta,\alpha,t}^{j}\left. \right\vert y_t^j\leq \mbox{VaR}_{\alpha,t}^j \right)=\beta.
\end{equation}
A tail level $\alpha$ equal to 1\% (2.5\% or 5\%) denotes a distress state of the world, while a level $\alpha$ equal to 50\% denotes a normal, or median, state. The financial institution $j$ contribution to systemic risk is defined by
\begin{equation}\label{eq:DCoVaR}
\Delta {\mbox{CoVaR}^{\leq}}_{\beta,\alpha,t}^j= {\mbox{CoVaR}^{\leq}}_{\beta,\alpha,t}^{j} - {\mbox{CoVaR}^{\leq}}_{\beta,0.5,t}^{j}.
\end{equation}
$\Delta {\mbox{CoVaR}^{\leq}}_{\beta,\alpha,t}^j$ captures the negative externality that financial institution $j$ imposes on the financial system. As described in the following sections, we estimate the CoVaR and the $\Delta \mbox{CoVaR}$ of main listed European banks over the time span from January 2007 to March 2020 by using different dependence structure.

The systemic risk measure estimation is divided in three steps. In the first step we estimate a univariate AR-GARCH model to the time series of log-returns and compute the VaR at the given tail level $\alpha$. In the second step we calibrate the dependence structure by applying different multivariate approaches. In the third step we estimate the systemic risk measure by means of a numerical integration and (or) a numerical inversion.

Let $S_t$ be the stock price process of a given financial institution and 
$$
y_t = \log\frac{S_t}{S_{t-1}}
$$ 
be its log-return process. We assume for the  log-return process an AR-GARCH model with GJR dynamics for the volatility, that is
\begin{equation}\label{eq:ArmaGarch}
\begin{array}{rcl}
y_t & = & ay_{t-1} + \sigma_{t}\varepsilon_{t} + c \\
\sigma_{t}^2 & = & \alpha_0 + \alpha_1\left(\left|\sigma_{t-1}\varepsilon_{t-1}\right|-\gamma\left(\sigma_{t-1}\varepsilon_{t-1}\right)\right)^2 + \beta_1\sigma_{t-1}^2
\end{array}
\end{equation}
where the innovation $\varepsilon_t$ are independent and identically distributed random variables with zero mean and unit variance. As observed in \cite{krbmf2011jbf}, the follow equality holds
\begin{equation}\label{eq:VaR}
\mbox{VaR}_{\alpha}(y_{t+1}) = ay_{t} + \sigma_{t}(\mbox{VaR}_{\alpha}(\varepsilon_{t+1})) + c.
\end{equation}
that in practice means that it is possible to compute the VaR on the basis of the quantile of a standardized random variable. It should be noted that a numerical inversion is usually needed to compute these quantiles.

After having estimated for each bank and for the system the univariate discrete-time dynamic volatility model defined in equation (\ref{eq:ArmaGarch}), we extract the innovations and estimate different dependence structures. We consider 
the multivariate normal tempered stable (MNTS) model, the multivariate generalized hyperbolic (MGH) model, as described in \cite{handbook2019bstff}, and the best copula function in terms of AIC among normal, $t$, BB1 and BB7 copulas, as described in \cite{deluca2018covar} and \cite{deluca2019var}. For the 
MNTS and MGH models
we estimate a 13-dimensional model by using an ad-hoc procedure implemented in R considering an expectation-maximization maximum-likelihood approach. Differently, the copulas are calibrated on bivariate time series (i.e. for each couple $j$, the couple $\varepsilon^{sys}$ and $\varepsilon^j$ is considered) and they are estimated through the {\it VineCopula} package of R. While the estimation of both the MNTS adn MGH is time-consuming from a computational point of view and for this reason we decide to run few estimation procedures as possible, thus instead of estimating a bivariate model we perform a 13-dimensional estimation, the estimation of copula functions in large dimension can be problematic from a numerical error perspective, thus instead of estimating a 13-dimensional model we perform a bivariate estimation.

After these two estimation steps, we forecast the one-day ahead volatility obtained from the estimated AR-GARCH parameters. While in the MNTS (MGH) model $\varepsilon^{sys}$ and $\varepsilon^j$ are assumed NTS (GH) distributed, in the copula model they are assumed skew-$t$ distributed. This means that to evaluate the univariate VaR for the each single bank $j$, while in the NTS (GH) case one needs to invert the cumulative distribution function obtained by means of the fast Fourier transform algorithm (\cite{handbook2019bstff}), in the skew-$t$ case the VaR can be directly obtained through the {\it qsstd} function of the {\it fGarch} package of R.

On the basis of the univariate and multivariate estimates, it is possible to evaluate the CoVaR for each financial institution $j$ and time $t$. While closed formula for the CoVaR are available under normal distributional assumptions (see \cite{bernard2012statistical}), for non-normal models a numerical integration procedure is needed (see \cite{girardi2013covar} and \cite{bernard2015conditional}). More in details, equation (\ref{eq:CoVaR}) can be written as
\begin{equation}\label{eq:a}
\frac{P\left(y_t^{sys}\leq {\mbox{CoVaR}^{\leq}}_{\beta,\alpha,t}^{j} , y_t^j\leq \mbox{VaR}_{\alpha,t}^j \right)}{P\left(y_t^i\leq \mbox{VaR}_{\alpha,t}^j \right)} =\beta
\end{equation}
and by the definition of VaR, it follows that
$$
P\left(y_t^{sys}\leq {\mbox{CoVaR}^{\leq}}_{\beta,\alpha,t}^{j}, y_t^i\leq \mbox{VaR}_{\alpha,t}^j \right) = \alpha\beta.
$$

In the MNTS and MGH cases, given the density $f_t^j$ of the bivariate random variable defined by $y_t^{sys}$ and $y_t^j$, then the following equality can be considered
\begin{equation}\label{eq:DensityIntegral}
\int_{-\infty}^{{\mbox{CoVaR}^{\leq}}_{\beta,\alpha,t}^{j}}\int_{-\infty}^{VaR_{\alpha,t}^j} f_t^j(x,y)dxdy = \alpha\beta.
\end{equation}
to obtain an estimate of CoVaR$^{\leq}$. The integral in equation (\ref{eq:DensityIntegral}) is evaluated by the numerical integration algorithm implemented in the {\it quadrature} function of the {\it mvQuad} package of R. Thus, the CoVaR estimate is obtained through the one dimensional root finding algorithm implemented in the {\it uniroot} function of the {\it stats} package of R. As observed by \cite{banulescu2019backtesting}, it is interesting to highlight that by slightly modifying the integral in equation (\ref{eq:DensityIntegral}) it is possible to obtain an estimate of the marginal expected shortfall.

It should be noted that, as in equation (\ref{eq:VaR}), the CoVaR can be computed by considering the density of the bivariate random variable defined by $\varepsilon_t^{sys}$ and $\varepsilon_t^j$ and then by performing a location and scale transformation based on the estimated AR-GARCH parameters.

In the copula function cases, as shown in \cite{bernard2012statistical}, if one considers the dependence structure between $y_t^{sys}$ and $y_t^j$, then the following equality can be considered
$$
P(y_t^{sys} \leq x, y_t^j \leq y) = C\left(F_{y_t^{sys}}(x), F_{y_t^j}(y)\right)
$$
and equation (\ref{eq:a}) can be rewritten as
$$
C\left( F_{y_t^{sys}}({CoVaR^{\leq}}_{\beta,\alpha,t}^{j}), F_{y_t^j}(VaR_{\alpha,t}^j)\right) = \alpha\beta
$$
where $C$ is a given copula function. By following the approach described in \cite{bernard2012statistical}, it is possible to obtain an estimate of the CoVaR by means of the inversion of the copula function. After having estimated the bivariate copula function the CoVaR estimate is obtained through the one dimensional root finding algorithm implemented in the {\it uniroot} function of the {\it stats} package of R.

As we will show in Section \ref{sec:Results} the CoVaR$^{\leq}$ based on these multivariate models is compared to the CoVaR$^{=}$ based on the multivariate normal GARCH model as defined by \cite{adrian2016covar} and empirically studied by \cite{bs2019covar}. 

\section{Data}\label{sec:Data}

Before starting with the empirical analysis, in this section we provide a description of the data used in this study. We obtained from Thomson Reuters Datastream daily dividend-adjusted closing prices from January 2002 to March 2020 for a sample of listed banks of the euro area belonging to the main or to the additional GSIB assessment sample for a total of twelve banks. These banks are: Deutsche Bank (D:DBK), Commerzbank (D:CBK),  Unicredit (I:UCG), Intesa (I:ISP), BNP Paribas (F:BNP), Soci\'et\'e G\'en\'erale (F:SGE), Cr\'edit Agricole (F:CRDA), BBVA (E:BBVA), Santander (E:SAN),  Banco de Sabadell (E:BSAB),  KBC Group (B:KC) and ING Bank (H:INGA). 
We refer to these European banks as GSIBs, even if for some of these banks there is no additional capital requirements. In this study the system is the Euro Stoxx 50 index. The time period in this analysis includes the high volatility period after the Lehman Brothers filing for Chapter 11 bankruptcy protection (September 15, 2008), the eurozone sovereign debt crisis, during which, in November 2011, the spread between the 10-year Italian BTP and the German Bund with the same maturity exceeded 500 basis points, the turmoil after the Italian political elections in 2018 and the recent financial market crash at the end of the first quarter of 2020.

The CoVaR ($\Delta \mbox{CoVaR}$) is estimated on the basis of the time series from the previous five years. For example, the CoVaR ($\Delta \mbox{CoVaR}$) for July 16, 2007 is estimated from the data for the period from July 16, 2002 to July 16, 2007. For each bank and each model we consider 3,389 estimations from January 2, 2007 to March 30, 2020.

\section{Empirical results}\label{sec:Results}

In this section we compare the different non-normal models with the multivariate normal one to which we refer to as MNormal. Additionally, we estimate both the CoVaR$^{\leq}$ and CoVaR$^{=}$ under this multivariate assumption. Recall that the CoVaR$^{=}$ under the MNormal assumption is the systemic risk measure originally proposed by \cite{adrian2016covar}.

As observed in Section \ref{sec:Methodology}, the first step is the estimation of the univariate autoregressive GARCH model with GJR dynamics for both system and banks log-returns. This step is performed through the {\it garchFit} function of the package {\it fGarch} of R. While for the multivariate models we consider a normal distributional assumption to extract the innovations, for the copula model we directly consider the skew-t distributional assumption. The former approach can be viewed as a quasi maximum-likelihood-estimation (QMLE) approach (see \cite{goode2015qmle}). Additionally, at each estimation step we verify if the autoregressive component is statistically significant: if it is not, we estimate the model without the autoregressive component.

After estimating the univariate dynamic volatility models, we extract the innovations and estimate the multivariate models (MNTS and MGH) and the copula ones. There are remarkable differences in terms of computational time between the two approaches. While for a one-step estimation of the four competitor bivariate copulas 0.9 seconds are needed, for a one-step estimation of the multivariate MNTS (MGH) model around 150 (10) seconds are needed. For each estimation day,  in the copula case we have 12 independent bivariate estimations and in the multivariate case we have a single multivariate estimation. This means that the computing time of the copula and MGH model is similar to the copula one, but the computing time of the MNTS model is 15 times larger. By considering that the overall computing time is around 24 hours for the faster models, to deal with the MNTS estimation problem we relied on 
an efficient R code making use of the packages {\it foreach} and {\it doParallel} and run it on a multi-core platform.

As far as the copula model is concerned, on the basis of the AIC criterion the $t$-copula is selected in most of the cases (around 88\% over the 40,668 bivariate estimation), in very few cases the normal copula (0.5\%) and in all other cases the BB1 (around 12\%). As observed above, it should be noted that only 0.9 seconds are needed to perform a one-step estimation of four bivariate copulas.

\begin{figure}
\begin{center}
\includegraphics[width=\columnwidth]{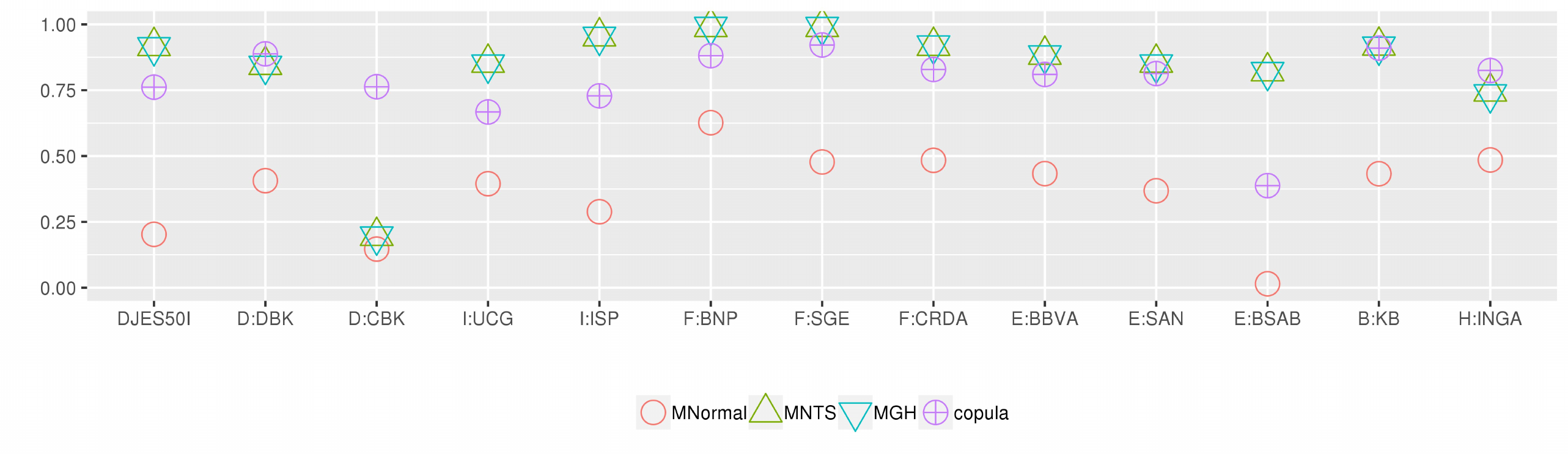}
\caption[]{\label{fig:KS}\footnotesize  For each margin and each model we report the average $p$-value of the KS test from January 2, 2007 to March 30, 2020.}
\end{center}
\end{figure}

For each model and each margin (i.e. each bank) Figure \ref{fig:KS} shows the average value over the entire estimation window from January 2, 2007 to March 30, 2020 of the $p$-value of the KS test. Even if the normal model has a satisfactory performance, in all cases the non-normal models are better than the normal one. The overall performance of the three non-normal models is comparable.

While for the copula model the computing time for VaR and CoVaR is instantaneous and the overall computing time is a few minutes, this is not the case for multivariate models. The time needed to estimate the CoVaR for the MNTS, the MGH and the MNormal model for a given day is around 8 minutes. For this reason also in this case we relied on a multi-core implementation. However, the VaR estimations are 60 times faster than the CoVaR ones. The main bottleneck is the numerical evaluation of the double integral in the MNTS case. It should be noted that the evaluation of the density function of a bivariate MNTS random variable is based on both a numerical integration and the FFT algorithm and, in the evaluation of the double integral, a large number of points is needed for the bidimensional grid in order to avoid numerical errors . 

\begin{figure}
\begin{center}
\includegraphics[width=\columnwidth]{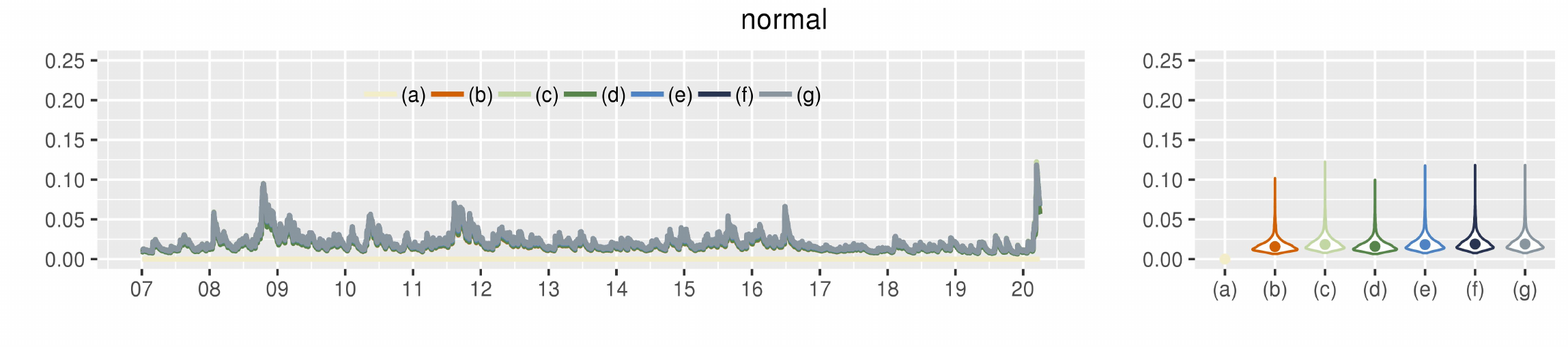}
\includegraphics[width=\columnwidth]{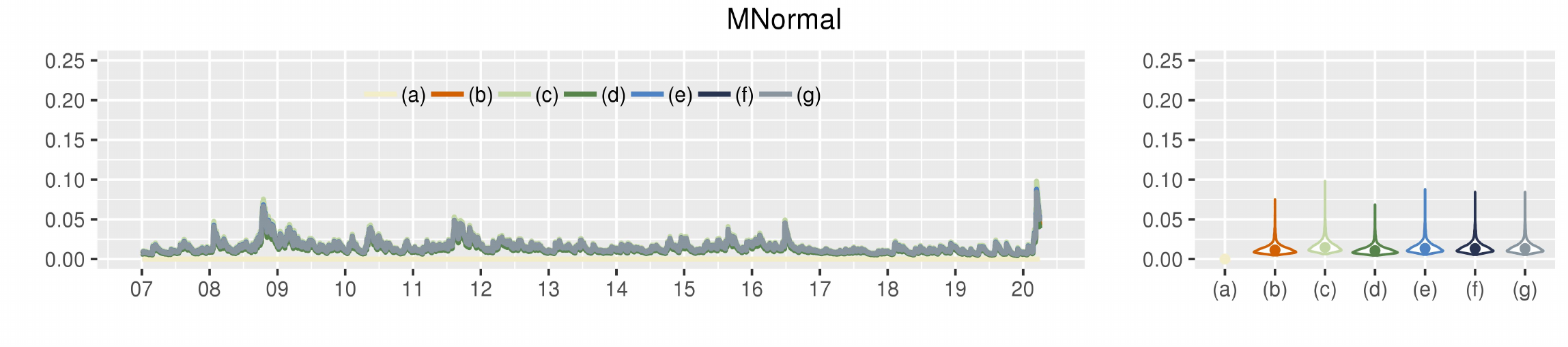}
\includegraphics[width=\columnwidth]{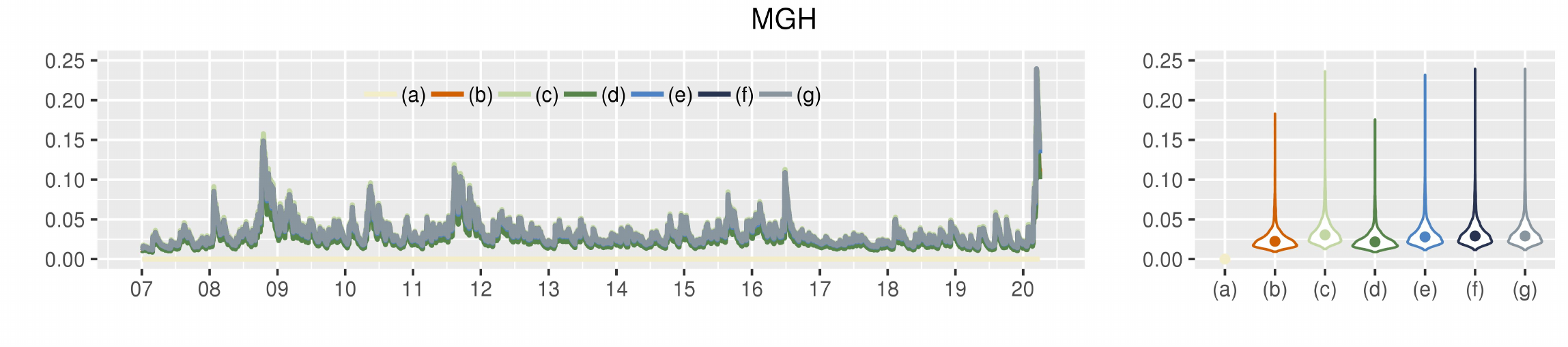}
\includegraphics[width=\columnwidth]{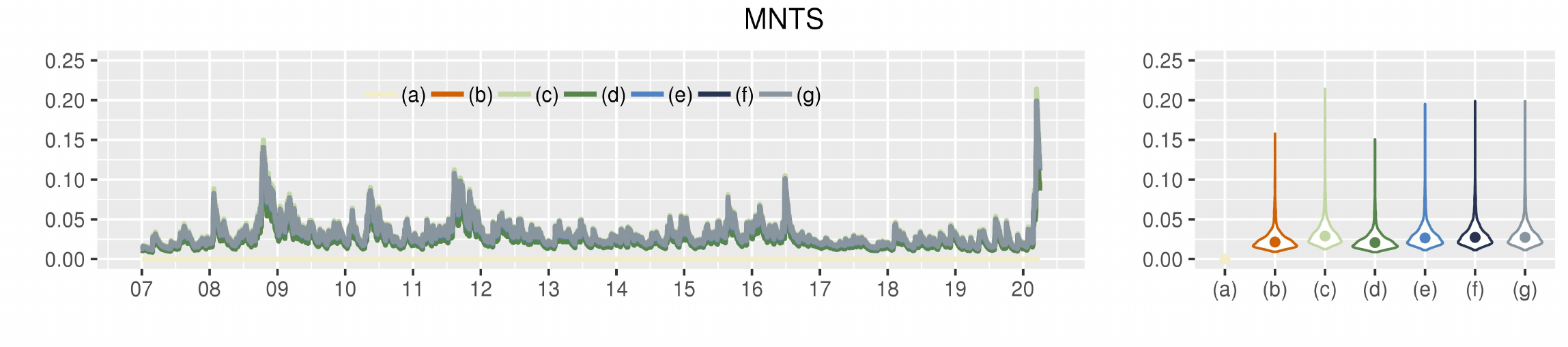}
\includegraphics[width=\columnwidth]{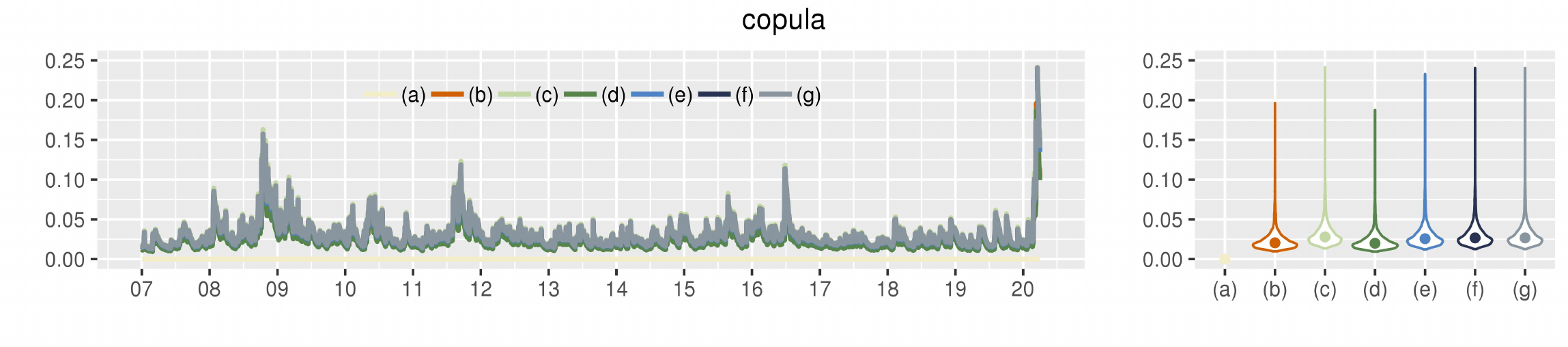}
\caption[]{\label{fig:CoVaR}\footnotesize  We report the average time series computed across all banks from January 2, 2007 to March 30, 2020 of the CoVaR$^{=}$ based on the multivariate normal model ({\it normal}) and the CoVaR$^{\leq}$ based on the MNormal, the MGH, the MNTS and the copula model for various level of $\alpha$ and $\beta$, that is (a) $\alpha = 0.5$ and $\beta = 0.025$, (b) $\alpha=\beta = 0.05$, (c) $\alpha = 0.05$ and $\beta = 0.025$, (d) $\alpha = 0.05$ and $\beta = 0.01$, (e) $\alpha = 0.025$ and $\beta = 0.05$, (f) $\alpha = \beta = 0.025$, and (g) $\alpha = 0.01$ and $\beta = 0.05$. We show the differences with respect to the (a) case. All values are changed in sign.}
\end{center}
\end{figure}

In order to show the differences between models, we show in Figure \ref{fig:CoVaR} the average time series computed across all banks from January 2, 2007 to March 30, 2020 of the CoVaR$^{=}$ based on the multivariate normal model and the CoVaR$^{\leq}$ based on  the MNormal, the MGH, the MNTS and the copula model for various level of $\alpha$ and $\beta$. In the MNormal and the three non-normal CoVaR$^{\leq}$ cases we report the differences with respect to the CoVaR$^{=}$. 

\begin{figure}
\begin{center}
\includegraphics[width=\columnwidth]{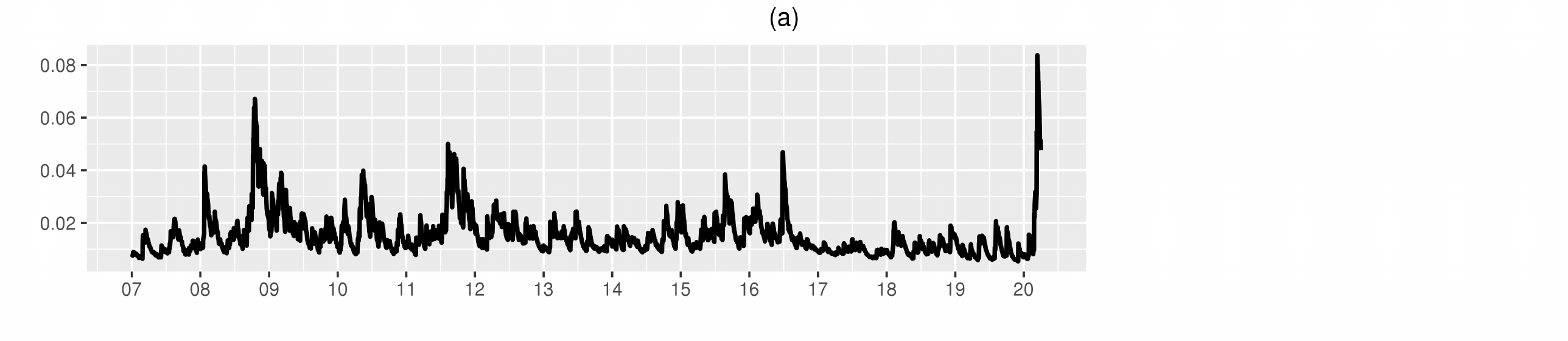}
\includegraphics[width=\columnwidth]{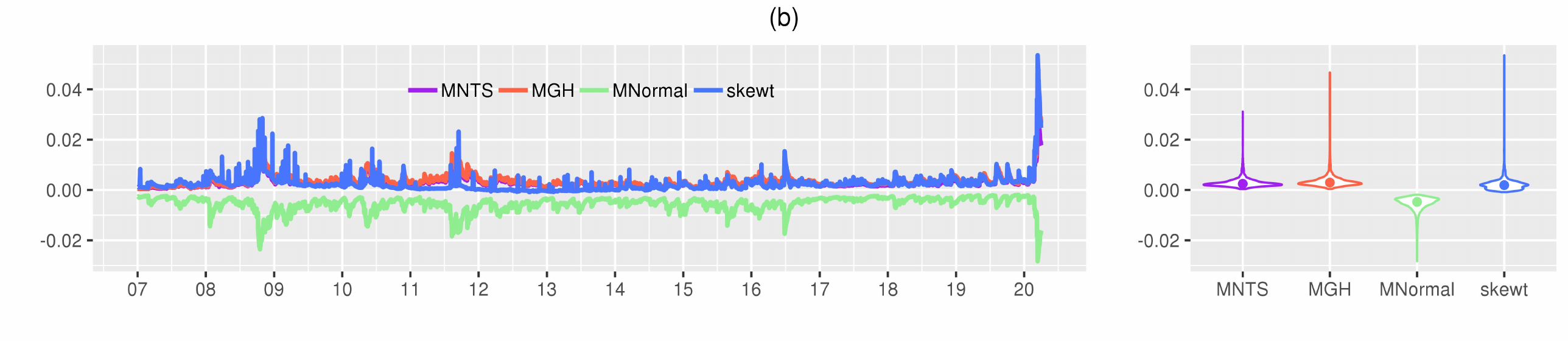}
\caption[]{\label{fig:DCoVaR_differences}\footnotesize  In the panel (a) we report the time series from January 2, 2007 to March 30, 2020 of the $\Delta \mbox{CoVaR}^{=}$ based on the multivariate normal GARCH model with GJR dynamics. In the panel (b) for each model we report for the same time window the time series and the violin plot of the difference between the $\Delta \mbox{CoVaR}^{\leq}$ and the above $\Delta \mbox{CoVaR}^{=}$. In all cases we consider $\alpha$ and $\beta$ equal to 0.05. All values are changed in sign.}
\end{center}
\end{figure}

In Figure \ref{fig:DCoVaR_differences} we report the time series of the average $\Delta \mbox{CoVaR}^{=}$  based on the multivariate normal model. We consider $\alpha$ and $\beta$ equal to 0.05 in equation (\ref{eq:DCoVaR}). The average is computed across all banks. Then, we compare the $\Delta CoVaR^{=}$ to the  $\Delta \mbox{CoVaR}^{\leq}$ evaluated under different distributional assumptions. In the MNormal case in panel (b) it appears a large difference between CoVaR$^{\leq}$ and CoVaR$^{=}$. This difference increases and changes in sign in non-normal cases. In all cases this difference varies over time. As shown in the violin plot on the right side of panel (b), it ranges from an average value of -0.55\% in MNormal case to 0.035\% in MGH case. The maximum distance is reached in March 2020 and it ranges from -2.8\% in the MNormal case to 5.3\% in the skew-t case. Recall that the violin plot is a method of drawing numeric data and combining a box plot with a kernel density plot.

\section{Backtesting}\label{sec:Backtest}

The definition (\ref{eq:CoVaR}) allows one to perform a two-steps backtesting. In both steps it is possible to follow the approach proposed in \cite{christoffersen2009backtesting}. First we conduct a preliminary VaR back-test by considering the entire observation window and defining a first hit sequence ($1$ if the loss of the financial institution on that day was larger than its predicted VaR level, and zero otherwise). Then we define a subset of observations on the basis on the distress of the financial institution $j$ (i.e. $y_t^i\leq \mbox{VaR}_{\alpha,t}^j$). Thus, by looking at this subset, we can backtest the CoVaR. More in details, we compare the CoVaR forecast with the ex-post loss of the financial system and define a second hit sequence which is $1$ if the loss of the financial system on that day is larger than its predicted CoVaR level, and zero otherwise.

\begin{figure}
\begin{center}
\includegraphics[width=\columnwidth]{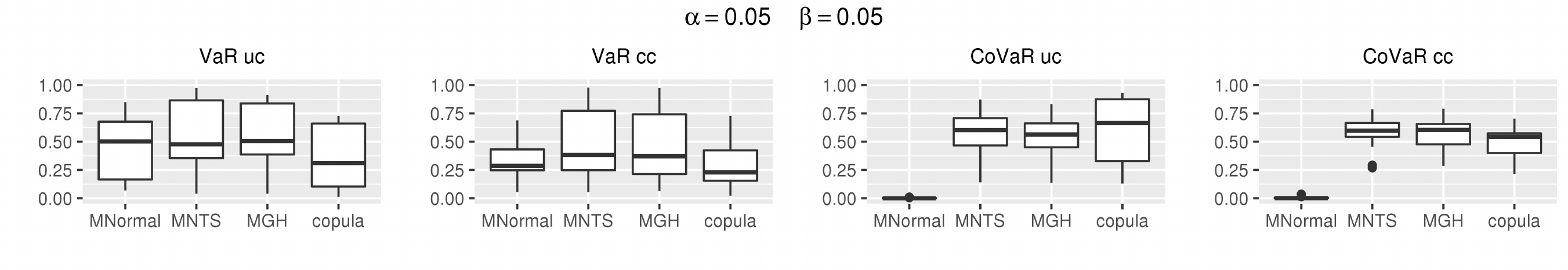}
\includegraphics[width=\columnwidth]{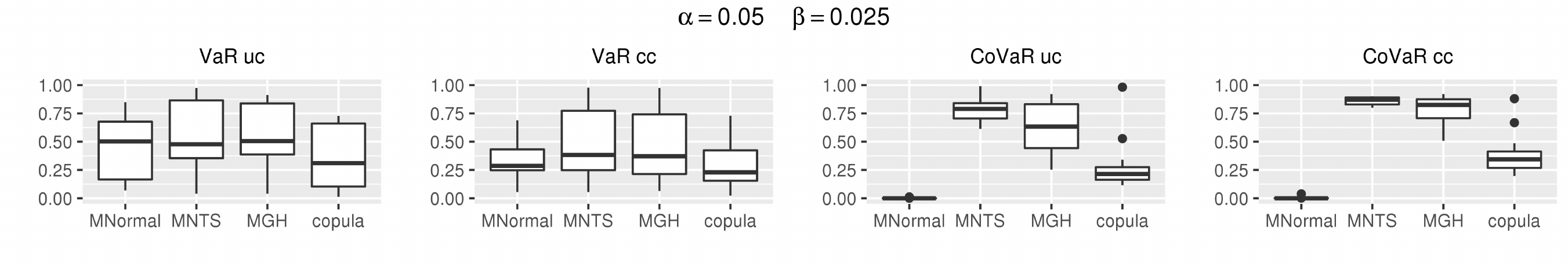}
\includegraphics[width=\columnwidth]{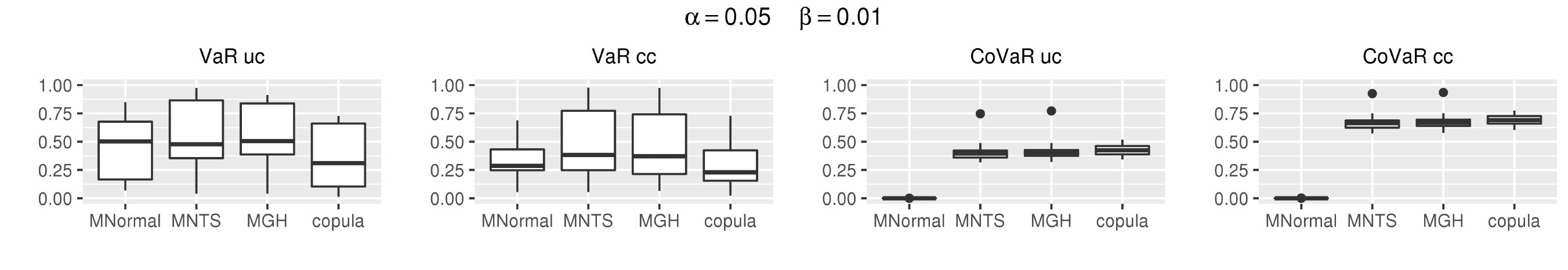}
\includegraphics[width=\columnwidth]{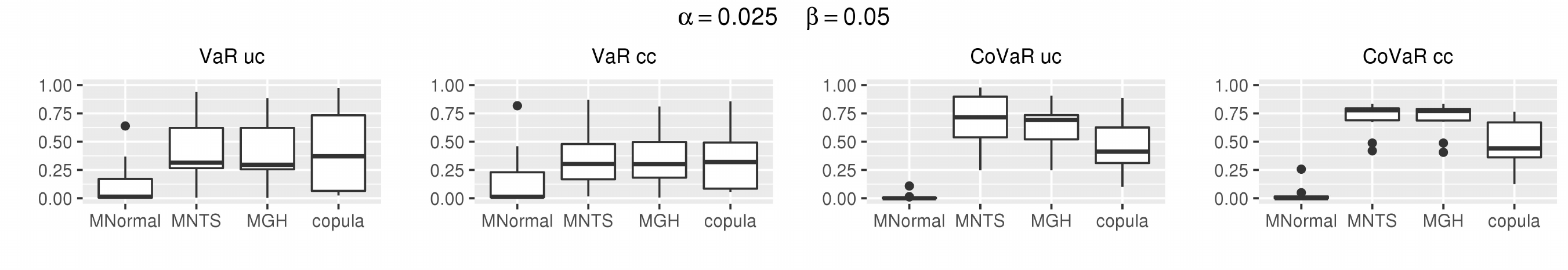}
\includegraphics[width=\columnwidth]{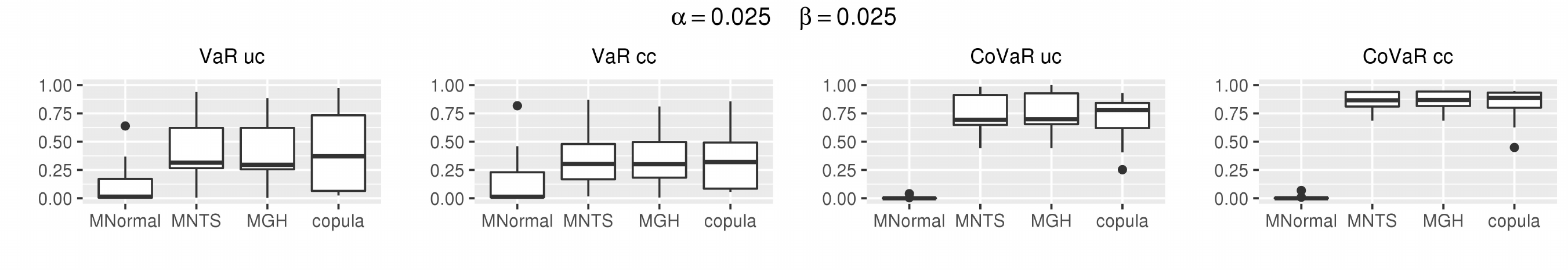}
\includegraphics[width=\columnwidth]{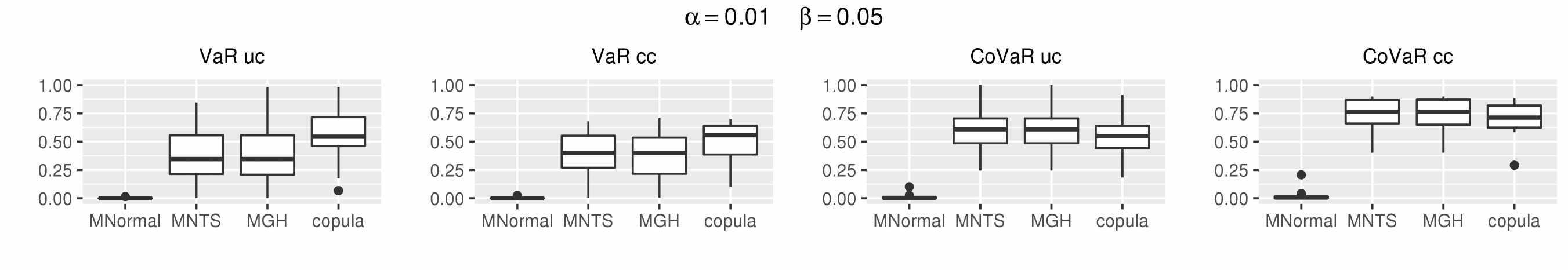}
\caption[]{\label{fig:Backtesting}\footnotesize  Boxplots across all banks of the $p$-values of the LR test of unconditional coverage (uc) and coverage and independence (cc) for both VaR and CoVaR for all analyzed models.}
\end{center}
\end{figure}

For evaluating the accuracy of forecasted VaR and CoVaR for the models analyzed in this paper, we perform the likelihood ratio (LR) tests proposed by \cite{christoffersen1998evaluating} and \cite{christoffersen2009backtesting}. The LR tests use the number of violations (i.e. the hit sequences defined above), where violations occur when the actual loss exceeds the estimated VaR (CoVaR). The LR test consists of three parts: (1) the LR test of unconditional coverage (LR$_{uc}$), which is the same as the proportion of failures test by \cite{kupiec1995techniques}, (2) the LR test of independence (LR$_{ind}$), and (3) the joint test of coverage and independence (LR$_{cc}$). In Figure \ref{fig:Backtesting} we report the $p$-values of the LR test of unconditional coverage (uc) and coverage and independence (cc) for both VaR and CoVaR for all analyzed models and various values of $\alpha$ and $\beta$. We do not report the $p$-values for the LR test of independence. The backtest for $\alpha=0.5$ and $\beta=0.05$ is not reported, even if it shows satisfactory results. By equation (\ref{eq:DCoVaR}), the tail level $\alpha=0.5$ is important to compute the $\Delta \mbox{CoVaR}$.
In order to backtest the CoVaR, a large number of observation is needed. Here we are considering 3,389 observations, this means that there are enough observations to backtest almost all tail levels $\alpha$, but it is not possible to backtest all meaningful tail levels $\beta$. For example, even if we are considering more than 18 years of daily data, it is not possible to study the CoVaR for $\alpha$ and $\beta$ equal to 0.01, because in this case the theoretical number of exceedances is less than one. 

We further consider the dynamic quantile (dq) test proposed by \cite{engle2004caviar}, that can be viewed as a more general formulation of the tests proposed above and two loss functions, that is the magnitude loss (LM) and the asymmetric magnitude loss (LA) function as defined by \cite{amendola2016evaluation}.
The farther the actual number of violations is from the expected one, the larger is the value of these loss functions. The LA function is built to penalize more heavily the models with a higher number of violations with respect to the number of expected violations.
The two functions are
\[
LM = 
\left\{
\begin{array}{lll}
1+(y_t-\mbox{VaR}_t)^2 & \hspace{0.5cm}\mbox{if}\hspace{0.5cm} & y_t < \mbox{VaR}_t \\
0 & \hspace{0.5cm}\mbox{if}\hspace{0.5cm}  & y_t \geq \mbox{VaR}_t 
\end{array}
\right.
\]
and
\[
LA = 
\left\{
\begin{array}{lll}
P(1+|y_t-\mbox{VaR}_t|) & \hspace{0.5cm}\mbox{if}\hspace{0.5cm} & y_t < \mbox{VaR}_t \\
|y_t-\mbox{VaR}_t| & \hspace{0.5cm}\mbox{if}\hspace{0.5cm}  & y_t \geq \mbox{VaR}_t 
\end{array}
\right.
\]
with $P=\exp((\hat{\alpha}-\alpha)/\alpha)$ if $\hat{\alpha}>\alpha$ and $P=1$
in the opposite case, where $\hat{\alpha}$ is the empirical coverage.
In Tables from \ref{tab:005} to \ref{tab:001} we report for each bank and for all $\alpha$ and $\beta$ considered in this study the $p$-values of the three tests (uc, cc and dq) and the values of the two loss functions (LM and LA).

As far as the VaR backtesting is concerned, the non-normal models (MNTS, MGH and copula) largely outperform the MNormal model, at least for the tail probability levels that are usually of interest (i.e. for $\alpha$ equal to 0.01 and 0.025). However, for $\alpha$ equal to 0.05, the MNormal model shows a better performance than the copula model, even if in the copula model case the null hypothesis is never rejected. This in practice means that capturing the volatility clustering effect is of paramount importance in evaluating risk measures, at least for the dataset analyzed in this study. Differently, in the CoVaR backtesting the MNormal model is always rejected and the non-normal model show always a satisfactory performance. By looking at the $p$-values of the three tests the MNTS and the MGH model seems slightly better in comparison with their competitor models. 

As far as the VaR estimates are concerned, the results in Tables  from \ref{tab:005} to \ref{tab:001} show that the LM function is smaller for the MGH and the MNTS models with respect to the copula model in 12 cases out of 12 when $\alpha = 0.05$, in 9 cases out of 12 when $\alpha = 0.025$, and in 8 cases out of 12 when $\alpha = 0.01$. Similar findings hold for the LA function, with the exception of the $\alpha = 0.01$ case. However, it should be noted that, on average, also the MNormal model show satisfactory results in terms of loss functions.

As far as the CoVaR estimates are concerned, the results in Tables  from \ref{tab:005} to \ref{tab:001} show that the LM function is smaller for the MGH and the MNTS models with respect to the copula model in 23 cases out of 36 when $\alpha = 0.05$, in 16 cases out of 24 when $\alpha = 0.025$, and in 4 cases out of 12 when $\alpha = 0.01$. Similar findings hold for the LA function. However, it should be noted that while, on average, there are not remarkable differences between the three non-normal models in terms of loss functions, the performance of the MNormal model is not satisfactory.

At least for the time-series and the values of $\alpha$ and $\beta$ analyzed in this study, the non-normal multivariate models (i.e. the MGH and the MNTS model) slightly outperform the copula from both a VaR and a CoVaR perspective. The MNormal approach is satisfactory only in the VaR case, but its performance is not good enough in the CoVaR backtesting.

\section{GSIBs indicators}\label{sec:GSIBs}

In this section, we compare the ranking provided by the $\Delta \mbox{CoVaR}$ with the ranking identified by the FSB for the GSIBs as defined in the \cite{fsb2011systemic} policy document and on its yearly updates. The FSB approach relies on firm-specific information on size, interconnectedness, substitutability, complexity, and cross-jurisdictional activity and it considers annual accounting and other data provided to regulators by financial institutions (see also \cite{bis2011systemic} and \cite{bis2013systemic}). 

More precisely, the GSIB framework analyze bank activities over 12 indicators. Each bank indicator is compared with the aggregate indicators of all banks in the specified sample (\cite{bis2014gsibscore}). These indicators are grouped into five categories (i.e. size, interconnectedness, substitutability, complexity and cross-jurisdictional activity). To calculate the score needed to determine the additional requirement, for each given indicator, the bank reported value for that indicator is divided by the corresponding sample total, and the resulting value is then expressed in basis points (bps). The final score is obtained as weighted average of the 12 indicators or as simple average of the five category scores (the average is rounded to the nearest whole basis point). Both specific bank  and  aggregated indicators are available on the FSB website.\footnote{ See 
} The score represents a bank activities as a percentage of the sample total and is used to determine the bank additional requirement. A higher score results in a higher requirement. On the basis on the indicators obtained from the FSB website we compute the score for a subsample of listed European banks for which these indicators are available (for Banco de Sabadell and KBC Group the indicators are not available). These scores are not equal to the official ones, because here we are considering only a subsample of banks.

By following the same approach implemented for the evaluation of the GSIBs indicators, for each model, we compute the annual average of the $\Delta \mbox{CoVaR}$ and rescale it in order to obtain the corresponding score expressed in basis points. We consider $\alpha$ and $\beta$ equal to 0.05. Both indicators and scores are reported  in Table \ref{tab:GSIBs}, where the sum of each column is by construction equal to 10,000 basis points (i.e. 100\%). The banks are ordered from the most to the less systemic according the GSIBs score. By looking at the values and at the colours of these scores, it appears evident that the variability across banks of the score defined on the basis of the $\Delta \mbox{CoVaR}$ is not high. Additionally, as shown in Figure \ref{fig:CoVaR}, there are large differences in terms of CoVaR between different models. The differences between models decrease when one evaluates the $\Delta \mbox{CoVaR}$ (see Figure \ref{fig:DCoVaR_differences}), and they decrease further when the annual average is computed. While, if one looks at daily estimates, the non-normal CoVaR is able to capture tail events, this seems not to be necessarily the case if one looks at annual averages. However, even if one looks at the annual maximum values, the overall results of the analysis are closed to those shown in Table \ref{tab:GSIBs} where the annual averages are considered.

In Table \ref{tab:GSIBs} we report the average indicators and scores over the years from 2013 to 2018. There is a large difference between the GSIBs score and the score based on the $\Delta \mbox{CoVaR}$, particularly for some banks (see also \cite{jokivuolle2018testing}). However, there are not remarkable differences in the $\Delta \mbox{CoVaR}$ scores computed under different distributional assumptions. 

Thus, we define the average relative percentage error (ARPE) with respect to the GSIBs score as
\begin{equation}\label{eq:ARPE}
ARPE = \frac{1}{nobs}\sum_{j,t} \frac{|score_{j,t}^{\mbox{GSIBs}} - score_{j,t}^{\Delta \mbox{CoVaR}}|}{score_{j,t}^{\mbox{GSIBs}}}
\end{equation}
where $nobs$ is the number of observations and $score_{j,t}^{\mbox{GSIBs}}$ and $score_{j,t}^{\Delta \mbox{CoVaR}}$ represent the GSIBs score and the score computed on the basis of the $\Delta \mbox{CoVaR}$ estimates. The ARPE from 2013 to 2018 across the ten banks in the sample is around 50\%.  As observed above, the GSIBs score is computed as
$$
score_{j,t}^{\mbox{GSIBs}} = \sum_i\omega_i Category_i
$$
where $i$ goes from 1 to 5 (size, interconnectedness, substitutability, complexity and cross-jurisdictional activity) and the weights $\omega_i$ are all equal to 0.2. 

Since the ARPE value is high, we try to modify the weights $\omega_i$ of the five categories in order to minimize the percentage difference in equation (\ref{eq:ARPE}), and we refer to them as {\it minimum distance weights}. By considering these weights, the ARPE with respect to the GSIBs score decreases to around 30\%.  

However, for all models only size and interconnectedness have minimum distance weights different from zero. The size weight range from 18\% for the MNTS model to 30\% for the normal model, and the interconnectedness ranges from 70\% for the normal model to 82\% for the MNTS model. It seems that the $\Delta \mbox{CoVaR}$ is correlated only with the size and interconnectedness. Conversely, it is not correlated with the other three categories.

Once we have these new weights, we can compute the corresponding score which is a sort of adjusted GSIBs score, as reported in Table \ref{tab:GSIBs}. This adjusted score can be view as a GSIBs score adjusted for the information coming from the stock market.

\section{Conclusions}\label{sec:Conclusions}

In this paper we compute the CoVaR and the $\Delta \mbox{CoVaR}$ under different distributional assumptions and different CoVaR definitions (i.e. CoVaR$^{=}$ and CoVaR$^{\leq}$). In order to backtest the CoVaR$^{\leq}$, we conduct an empirical test over more than 18 years of daily stock log-returns. We calibrate three different models allowing for volatility clustering, heavy tails and non-linear dependence. These multivariate non-normal models outperform the normal one in term of log-returns fitting, VaR and CoVaR backtesting for all tail levels analyzed in this study. The backtesting exercise shows that the performance of the MGH (MNTS) model is slightly better compared to the copula model, however this latter approach is very promising from a computational point of view. Finally, we compare the score obtained through the $\Delta \mbox{CoVaR}$ measure with the GSIBs score defined by the FSB and introduce a score adjusted for the information coming from the stock markets.



\begin{sidewaystable}
\centering
\begin{scriptsize}
\begin{tabular}{ll|ccccc|ccccc|ccccc|ccccc}
\toprule

			&				&	\multicolumn{5}{c}{MNormal}									&	\multicolumn{5}{c}{MGH}									&	\multicolumn{5}{c}{MNTS}									&	\multicolumn{5}{c}{Copula}									\\
			&				&	uc	&	cc	&	dq	&	LM	&	LA	&	uc	&	cc	&	dq	&	LM	&	LA	&	uc	&	cc	&	dq	&	LM	&	LA	&	uc	&	cc	&	dq	&	LM	&	LA	\\

\midrule
\rowcolor{gray!10}
	&	$	VaR \quad (\alpha =  0.05)	$	&	0.45	&	0.66	&	0.08	&	0.05	&	0.09	&	0.84	&	0.97	&	0.25	&	0.05	&	0.09	&	0.85	&	0.98	&	0.15	&	0.05	&	0.09	&	0.72	&	0.73	&	0.06	&	0.05	&	0.09	\\
\rowcolor{gray!10}
			&	$	CoVaR \quad (\beta =  0.05)	$	&	0.00	&	0.01	&	0.02	&	0.11	&	0.13	&	0.56	&	0.66	&	0.93	&	0.04	&	0.07	&	0.62	&	0.68	&	0.93	&	0.04	&	0.07	&	0.66	&	0.52	&	0.88	&	0.06	&	0.08	\\
\rowcolor{gray!10}
			&	$	CoVaR \quad (\beta =  0.025)	$	&	0.00	&	0.00	&	0.00	&	0.08	&	0.11	&	0.88	&	0.92	&	1.00	&	0.02	&	0.06	&	0.69	&	0.82	&	1.00	&	0.03	&	0.07	&	0.24	&	0.39	&	0.80	&	0.04	&	0.07	\\
\rowcolor{gray!10}
\multirow{-4}{*}{D:DBK}			&	$	CoVaR \quad (\beta =  0.01)	$	&	0.00	&	0.00	&	0.00	&	0.06	&	0.15	&	0.37	&	0.65	&	1.00	&	0.02	&	0.06	&	0.35	&	0.63	&	1.00	&	0.02	&	0.06	&	0.38	&	0.66	&	1.00	&	0.02	&	0.06	\\
\multirow{4}{*}{D:CBK}	&	$	VaR \quad (\alpha =  0.05)	$	&	0.73	&	0.36	&	0.58	&	0.05	&	0.09	&	0.41	&	0.16	&	0.08	&	0.05	&	0.10	&	0.50	&	0.16	&	0.07	&	0.05	&	0.10	&	0.23	&	0.46	&	0.27	&	0.05	&	0.10	\\
			&	$	CoVaR \quad (\beta =  0.05)	$	&	0.00	&	0.00	&	0.00	&	0.12	&	0.14	&	0.51	&	0.42	&	0.54	&	0.06	&	0.09	&	0.31	&	0.27	&	0.26	&	0.07	&	0.10	&	0.93	&	0.66	&	0.88	&	0.05	&	0.08	\\
			&	$	CoVaR \quad (\beta =  0.025)	$	&	0.00	&	0.00	&	0.00	&	0.08	&	0.12	&	0.81	&	0.87	&	1.00	&	0.03	&	0.06	&	0.80	&	0.86	&	1.00	&	0.03	&	0.06	&	0.15	&	0.26	&	0.61	&	0.04	&	0.08	\\
			&	$	CoVaR \quad (\beta =  0.01)	$	&	0.00	&	0.00	&	0.00	&	0.07	&	0.51	&	0.41	&	0.69	&	1.00	&	0.02	&	0.06	&	0.40	&	0.68	&	1.00	&	0.02	&	0.06	&	0.44	&	0.71	&	1.00	&	0.02	&	0.06	\\
\rowcolor{gray!10}
	&	$	VaR \quad (\alpha =  0.05)	$	&	0.17	&	0.39	&	0.65	&	0.06	&	0.10	&	0.41	&	0.70	&	0.84	&	0.05	&	0.10	&	0.45	&	0.75	&	0.87	&	0.05	&	0.10	&	0.08	&	0.20	&	0.41	&	0.06	&	0.10	\\
\rowcolor{gray!10}
			&	$	CoVaR \quad (\beta =  0.05)	$	&	0.00	&	0.01	&	0.12	&	0.10	&	0.12	&	0.14	&	0.29	&	0.87	&	0.03	&	0.06	&	0.14	&	0.29	&	0.87	&	0.03	&	0.06	&	0.90	&	0.57	&	0.74	&	0.05	&	0.08	\\
\rowcolor{gray!10}
			&	$	CoVaR \quad (\beta =  0.025)	$	&	0.00	&	0.00	&	0.00	&	0.07	&	0.10	&	0.45	&	0.71	&	0.98	&	0.02	&	0.05	&	0.82	&	0.89	&	1.00	&	0.02	&	0.06	&	0.34	&	0.49	&	0.92	&	0.04	&	0.07	\\
\rowcolor{gray!10}
\multirow{-4}{*}{I:UCG}			&	$	CoVaR \quad (\beta =  0.01)	$	&	0.00	&	0.00	&	0.00	&	0.05	&	0.10	&	0.41	&	0.68	&	1.00	&	0.02	&	0.06	&	0.41	&	0.67	&	1.00	&	0.02	&	0.06	&	0.47	&	0.73	&	1.00	&	0.02	&	0.06	\\
\multirow{4}{*}{I:ISP}	&	$	VaR \quad (\alpha =  0.05)	$	&	0.66	&	0.69	&	0.53	&	0.05	&	0.09	&	0.91	&	0.88	&	0.56	&	0.05	&	0.09	&	0.91	&	0.88	&	0.56	&	0.05	&	0.09	&	0.37	&	0.42	&	0.38	&	0.05	&	0.09	\\
			&	$	CoVaR \quad (\beta =  0.05)	$	&	0.00	&	0.00	&	0.00	&	0.12	&	0.14	&	0.58	&	0.45	&	0.63	&	0.06	&	0.09	&	0.58	&	0.45	&	0.64	&	0.06	&	0.09	&	0.34	&	0.27	&	0.49	&	0.07	&	0.09	\\
			&	$	CoVaR \quad (\beta =  0.025)	$	&	0.00	&	0.00	&	0.00	&	0.10	&	0.14	&	0.92	&	0.90	&	1.00	&	0.02	&	0.06	&	0.92	&	0.90	&	1.00	&	0.02	&	0.06	&	0.13	&	0.23	&	0.31	&	0.04	&	0.08	\\
			&	$	CoVaR \quad (\beta =  0.01)	$	&	0.00	&	0.00	&	0.00	&	0.07	&	0.56	&	0.36	&	0.62	&	1.00	&	0.02	&	0.06	&	0.36	&	0.62	&	1.00	&	0.02	&	0.06	&	0.42	&	0.68	&	1.00	&	0.02	&	0.06	\\
\rowcolor{gray!10}
	&	$	VaR \quad (\alpha =  0.05)	$	&	0.66	&	0.55	&	0.91	&	0.05	&	0.09	&	0.84	&	0.89	&	0.99	&	0.05	&	0.09	&	0.97	&	0.86	&	0.99	&	0.05	&	0.09	&	0.66	&	0.38	&	0.87	&	0.05	&	0.09	\\
\rowcolor{gray!10}
		&	$	CoVaR \quad (\beta =  0.05)	$	&	0.00	&	0.00	&	0.01	&	0.12	&	0.14	&	0.83	&	0.66	&	0.91	&	0.05	&	0.08	&	0.87	&	0.66	&	0.91	&	0.05	&	0.08	&	0.45	&	0.70	&	0.71	&	0.06	&	0.09	\\
\rowcolor{gray!10}
			&	$	CoVaR \quad (\beta =  0.025)	$	&	0.00	&	0.00	&	0.00	&	0.10	&	0.13	&	0.50	&	0.76	&	0.99	&	0.02	&	0.05	&	0.91	&	0.90	&	1.00	&	0.02	&	0.06	&	0.12	&	0.20	&	0.52	&	0.05	&	0.08	\\
\rowcolor{gray!10}
\multirow{-4}{*}{F:BNP}			&	$	CoVaR \quad (\beta =  0.01)	$	&	0.00	&	0.00	&	0.00	&	0.07	&	0.53	&	0.37	&	0.64	&	1.00	&	0.02	&	0.06	&	0.36	&	0.62	&	1.00	&	0.02	&	0.06	&	0.39	&	0.65	&	1.00	&	0.02	&	0.06	\\
\multirow{4}{*}{F:SGE}	&	$	VaR \quad (\alpha =  0.05)	$	&	0.41	&	0.05	&	0.21	&	0.05	&	0.09	&	0.45	&	0.06	&	0.29	&	0.05	&	0.09	&	0.41	&	0.05	&	0.25	&	0.05	&	0.09	&	0.73	&	0.12	&	0.47	&	0.05	&	0.09	\\
			&	$	CoVaR \quad (\beta =  0.05)	$	&	0.00	&	0.00	&	0.00	&	0.12	&	0.14	&	0.71	&	0.68	&	0.93	&	0.04	&	0.07	&	0.72	&	0.68	&	0.93	&	0.04	&	0.07	&	0.79	&	0.57	&	0.92	&	0.05	&	0.08	\\
			&	$	CoVaR \quad (\beta =  0.025)	$	&	0.00	&	0.00	&	0.00	&	0.09	&	0.13	&	0.60	&	0.82	&	0.99	&	0.02	&	0.05	&	0.99	&	0.90	&	1.00	&	0.03	&	0.06	&	0.19	&	0.31	&	0.66	&	0.04	&	0.08	\\
			&	$	CoVaR \quad (\beta =  0.01)	$	&	0.00	&	0.00	&	0.00	&	0.07	&	0.37	&	0.32	&	0.58	&	1.00	&	0.02	&	0.06	&	0.32	&	0.57	&	1.00	&	0.02	&	0.06	&	0.34	&	0.60	&	1.00	&	0.02	&	0.06	\\
\rowcolor{gray!10}
\multirow{4}{*}{F:CRDA}	&	$	VaR \quad (\alpha =  0.05)	$	&	0.85	&	0.26	&	0.03	&	0.05	&	0.09	&	0.84	&	0.35	&	0.04	&	0.05	&	0.09	&	0.97	&	0.30	&	0.04	&	0.05	&	0.09	&	0.55	&	0.25	&	0.15	&	0.05	&	0.09	\\
\rowcolor{gray!10}
			&	$	CoVaR \quad (\beta =  0.05)	$	&	0.00	&	0.00	&	0.00	&	0.13	&	0.15	&	0.56	&	0.63	&	0.89	&	0.04	&	0.07	&	0.85	&	0.59	&	0.85	&	0.05	&	0.08	&	0.30	&	0.24	&	0.53	&	0.07	&	0.09	\\
\rowcolor{gray!10}
			&	$	CoVaR \quad (\beta =  0.025)	$	&	0.00	&	0.00	&	0.00	&	0.10	&	0.15	&	0.88	&	0.90	&	0.99	&	0.02	&	0.06	&	0.71	&	0.80	&	0.97	&	0.03	&	0.07	&	0.25	&	0.39	&	0.69	&	0.04	&	0.07	\\
\rowcolor{gray!10}
\multirow{-4}{*}{F:CRDA}			&	$	CoVaR \quad (\beta =  0.01)	$	&	0.00	&	0.00	&	0.00	&	0.08	&	0.77	&	0.37	&	0.64	&	0.99	&	0.02	&	0.07	&	0.36	&	0.62	&	0.99	&	0.02	&	0.06	&	0.40	&	0.66	&	0.99	&	0.02	&	0.06	\\
\multirow{4}{*}{E:BBVA}	&	$	VaR \quad (\alpha =  0.05)	$	&	0.13	&	0.31	&	0.11	&	0.06	&	0.09	&	0.33	&	0.51	&	0.21	&	0.05	&	0.09	&	0.33	&	0.51	&	0.20	&	0.05	&	0.09	&	0.06	&	0.16	&	0.03	&	0.06	&	0.09	\\
			&	$	CoVaR \quad (\beta =  0.05)	$	&	0.00	&	0.00	&	0.00	&	0.12	&	0.14	&	0.46	&	0.57	&	0.94	&	0.04	&	0.07	&	0.46	&	0.57	&	0.94	&	0.04	&	0.06	&	0.92	&	0.58	&	0.93	&	0.05	&	0.08	\\
			&	$	CoVaR \quad (\beta =  0.025)	$	&	0.00	&	0.00	&	0.00	&	0.10	&	0.13	&	0.43	&	0.70	&	0.98	&	0.02	&	0.05	&	0.79	&	0.88	&	1.00	&	0.02	&	0.05	&	0.18	&	0.29	&	0.72	&	0.04	&	0.07	\\
			&	$	CoVaR \quad (\beta =  0.01)	$	&	0.00	&	0.00	&	0.00	&	0.07	&	0.32	&	0.42	&	0.69	&	1.00	&	0.02	&	0.06	&	0.42	&	0.69	&	1.00	&	0.02	&	0.06	&	0.48	&	0.74	&	1.00	&	0.02	&	0.05	\\
\rowcolor{gray!10}
	&	$	VaR \quad (\alpha =  0.05)	$	&	0.07	&	0.12	&	0.10	&	0.06	&	0.09	&	0.20	&	0.23	&	0.22	&	0.05	&	0.09	&	0.33	&	0.38	&	0.29	&	0.05	&	0.09	&	0.11	&	0.19	&	0.17	&	0.06	&	0.09	\\
\rowcolor{gray!10}
			&	$	CoVaR \quad (\beta =  0.05)	$	&	0.00	&	0.00	&	0.01	&	0.11	&	0.13	&	0.65	&	0.63	&	0.94	&	0.04	&	0.07	&	0.70	&	0.64	&	0.94	&	0.04	&	0.07	&	0.87	&	0.56	&	0.90	&	0.05	&	0.08	\\
\rowcolor{gray!10}
			&	$	CoVaR \quad (\beta =  0.025)	$	&	0.00	&	0.00	&	0.00	&	0.09	&	0.12	&	0.41	&	0.68	&	0.98	&	0.02	&	0.05	&	0.79	&	0.88	&	1.00	&	0.02	&	0.05	&	0.17	&	0.27	&	0.71	&	0.04	&	0.07	\\
\rowcolor{gray!10}
\multirow{-4}{*}{E:SAN}			&	$	CoVaR \quad (\beta =  0.01)	$	&	0.00	&	0.00	&	0.00	&	0.07	&	0.28	&	0.44	&	0.71	&	1.00	&	0.02	&	0.06	&	0.42	&	0.69	&	1.00	&	0.02	&	0.06	&	0.46	&	0.72	&	1.00	&	0.02	&	0.05	\\
\multirow{4}{*}{E:BSAB}	&	$	VaR \quad (\alpha =  0.05)	$	&	0.15	&	0.25	&	0.02	&	0.06	&	0.09	&	0.04	&	0.11	&	0.01	&	0.06	&	0.09	&	0.04	&	0.11	&	0.01	&	0.06	&	0.09	&	0.01	&	0.02	&	0.00	&	0.06	&	0.09	\\
			&	$	CoVaR \quad (\beta =  0.05)	$	&	0.01	&	0.04	&	0.07	&	0.10	&	0.12	&	0.33	&	0.48	&	0.94	&	0.04	&	0.07	&	0.54	&	0.59	&	0.96	&	0.04	&	0.07	&	0.29	&	0.44	&	0.94	&	0.03	&	0.06	\\
			&	$	CoVaR \quad (\beta =  0.025)	$	&	0.01	&	0.04	&	0.00	&	0.06	&	0.08	&	0.67	&	0.84	&	0.99	&	0.02	&	0.06	&	0.67	&	0.84	&	0.99	&	0.02	&	0.06	&	0.98	&	0.88	&	1.00	&	0.02	&	0.06	\\
			&	$	CoVaR \quad (\beta =  0.01)	$	&	0.00	&	0.00	&	0.00	&	0.04	&	0.08	&	0.49	&	0.75	&	1.00	&	0.02	&	0.06	&	0.49	&	0.75	&	1.00	&	0.02	&	0.06	&	0.52	&	0.78	&	1.00	&	0.01	&	0.06	\\
\rowcolor{gray!10}
	&	$	VaR \quad (\alpha =  0.05)	$	&	0.55	&	0.25	&	0.03	&	0.05	&	0.10	&	0.61	&	0.39	&	0.04	&	0.05	&	0.10	&	0.66	&	0.38	&	0.04	&	0.05	&	0.10	&	0.26	&	0.42	&	0.12	&	0.05	&	0.10	\\
\rowcolor{gray!10}
			&	$	CoVaR \quad (\beta =  0.05)	$	&	0.00	&	0.00	&	0.00	&	0.15	&	0.18	&	0.68	&	0.79	&	0.84	&	0.06	&	0.09	&	0.67	&	0.79	&	0.84	&	0.06	&	0.09	&	0.13	&	0.22	&	0.16	&	0.08	&	0.11	\\
\rowcolor{gray!10}
			&	$	CoVaR \quad (\beta =  0.025)	$	&	0.00	&	0.00	&	0.00	&	0.10	&	0.15	&	0.78	&	0.83	&	1.00	&	0.03	&	0.07	&	0.77	&	0.83	&	1.00	&	0.03	&	0.07	&	0.53	&	0.67	&	0.97	&	0.03	&	0.07	\\
\rowcolor{gray!10}
\multirow{-4}{*}{B:KB}			&	$	CoVaR \quad (\beta =  0.01)	$	&	0.00	&	0.00	&	0.00	&	0.07	&	0.33	&	0.39	&	0.66	&	1.00	&	0.02	&	0.07	&	0.39	&	0.65	&	1.00	&	0.02	&	0.06	&	0.43	&	0.70	&	1.00	&	0.02	&	0.06	\\
\multirow{4}{*}{H:INGA}	&	$	VaR \quad (\alpha =  0.05)	$	&	0.78	&	0.23	&	0.68	&	0.05	&	0.09	&	0.56	&	0.27	&	0.66	&	0.05	&	0.09	&	0.36	&	0.28	&	0.58	&	0.05	&	0.09	&	0.66	&	0.14	&	0.33	&	0.05	&	0.09	\\
			&	$	CoVaR \quad (\beta =  0.05)	$	&	0.00	&	0.00	&	0.00	&	0.13	&	0.15	&	0.43	&	0.58	&	0.96	&	0.04	&	0.06	&	0.47	&	0.61	&	0.96	&	0.04	&	0.06	&	0.67	&	0.50	&	0.91	&	0.06	&	0.08	\\
			&	$	CoVaR \quad (\beta =  0.025)	$	&	0.00	&	0.00	&	0.00	&	0.10	&	0.15	&	0.25	&	0.51	&	0.99	&	0.01	&	0.05	&	0.61	&	0.83	&	1.00	&	0.02	&	0.05	&	0.24	&	0.38	&	0.80	&	0.04	&	0.07	\\
			&	$	CoVaR \quad (\beta =  0.01)	$	&	0.00	&	0.00	&	0.00	&	0.08	&	0.57	&	0.77	&	0.94	&	1.00	&	0.01	&	0.06	&	0.75	&	0.93	&	1.00	&	0.01	&	0.05	&	0.39	&	0.65	&	1.00	&	0.02	&	0.06	\\

\bottomrule
			
\end{tabular}
\caption[]{\label{tab:005}\footnotesize $P$-values of the LR test of unconditional coverage (uc), coverage and independence (cc) and dynamic quantile tests (dq) for both $VaR$ and $CoVaR$  with $\alpha = 0.05$ for all analyzed models and banks. The values of the magnitude (LM) and of the asymmetric loss (LA) functions are also reported. }
\end{scriptsize}
\end{sidewaystable}


\begin{sidewaystable}
\begin{center}
\begin{scriptsize}
\begin{tabular}{ll|ccccc|ccccc|ccccc|ccccc}
\toprule

			&				&	\multicolumn{5}{c}{MNormal}									&	\multicolumn{5}{c}{\textbf{MGH}}									&	\multicolumn{5}{c}{MNTS}									&	\multicolumn{5}{c}{Copula}									\\
			&				&	uc	&	cc	&	dq	&	LM	&	LA	&	uc	&	cc	&	dq	&	LM	&	LA	&	uc	&	cc	&	dq	&	LM	&	LA	&	uc	&	cc	&	dq	&	LM	&	LA	\\

\midrule
\rowcolor{gray!10}
	&	$	VaR \quad (\alpha =  0.025)	$	&	0.01	&	0.01	&	0.04	&	0.03	&	0.08	&	0.27	&	0.49	&	0.73	&	0.03	&	0.08	&	0.27	&	0.49	&	0.73	&	0.03	&	0.08	&	0.27	&	0.49	&	0.51	&	0.03	&	0.08	\\
\rowcolor{gray!10}
			&	$	CoVaR \quad (\beta =  0.05)	$	&	0.01	&	0.05	&	0.01	&	0.11	&	0.13	&	0.91	&	0.79	&	0.99	&	0.05	&	0.08	&	0.91	&	0.79	&	0.99	&	0.05	&	0.08	&	0.32	&	0.38	&	0.74	&	0.07	&	0.10	\\
\rowcolor{gray!10}
\multirow{-3}{*}{D:DBK}			&	$	CoVaR \quad (\beta =  0.025)	$	&	0.00	&	0.00	&	0.00	&	0.10	&	0.14	&	0.69	&	0.87	&	1.00	&	0.03	&	0.07	&	0.69	&	0.87	&	1.00	&	0.03	&	0.07	&	0.69	&	0.87	&	1.00	&	0.03	&	0.07	\\
\multirow{3}{*}{D:CBK}	&	$	VaR \quad (\alpha =  0.025)	$	&	0.64	&	0.82	&	0.51	&	0.03	&	0.08	&	0.53	&	0.81	&	0.69	&	0.02	&	0.08	&	0.60	&	0.87	&	0.70	&	0.02	&	0.08	&	0.89	&	0.86	&	0.37	&	0.03	&	0.08	\\
			&	$	CoVaR \quad (\beta =  0.05)	$	&	0.00	&	0.00	&	0.00	&	0.15	&	0.18	&	0.32	&	0.41	&	0.85	&	0.08	&	0.11	&	0.34	&	0.42	&	0.86	&	0.08	&	0.11	&	0.10	&	0.13	&	0.32	&	0.09	&	0.12	\\
			&	$	CoVaR \quad (\beta =  0.025)	$	&	0.00	&	0.00	&	0.00	&	0.13	&	0.29	&	0.49	&	0.73	&	1.00	&	0.04	&	0.08	&	0.50	&	0.74	&	1.00	&	0.04	&	0.07	&	0.25	&	0.45	&	0.98	&	0.05	&	0.08	\\
\rowcolor{gray!10}
	&	$	VaR \quad (\alpha =  0.025)	$	&	0.08	&	0.19	&	0.35	&	0.03	&	0.09	&	0.28	&	0.16	&	0.16	&	0.02	&	0.08	&	0.28	&	0.16	&	0.16	&	0.02	&	0.08	&	0.68	&	0.38	&	0.51	&	0.02	&	0.08	\\
\rowcolor{gray!10}
			&	$	CoVaR \quad (\beta =  0.05)	$	&	0.00	&	0.01	&	0.00	&	0.13	&	0.15	&	0.90	&	0.84	&	0.99	&	0.05	&	0.08	&	0.90	&	0.84	&	1.00	&	0.05	&	0.08	&	0.35	&	0.43	&	0.75	&	0.07	&	0.10	\\
\rowcolor{gray!10}
\multirow{-3}{*}{I:UCG}			&	$	CoVaR \quad (\beta =  0.025)	$	&	0.00	&	0.00	&	0.00	&	0.09	&	0.12	&	0.44	&	0.69	&	1.00	&	0.04	&	0.08	&	0.44	&	0.69	&	1.00	&	0.04	&	0.08	&	0.52	&	0.75	&	1.00	&	0.04	&	0.07	\\
\multirow{3}{*}{I:ISP}	&	$	VaR \quad (\alpha =  0.025)	$	&	0.01	&	0.01	&	0.14	&	0.03	&	0.08	&	0.22	&	0.22	&	0.64	&	0.03	&	0.08	&	0.27	&	0.26	&	0.69	&	0.03	&	0.08	&	0.06	&	0.07	&	0.27	&	0.03	&	0.08	\\
			&	$	CoVaR \quad (\beta =  0.05)	$	&	0.00	&	0.00	&	0.00	&	0.14	&	0.17	&	0.70	&	0.81	&	0.99	&	0.04	&	0.07	&	0.91	&	0.79	&	0.99	&	0.05	&	0.08	&	0.41	&	0.46	&	0.80	&	0.07	&	0.10	\\
			&	$	CoVaR \quad (\beta =  0.025)	$	&	0.00	&	0.00	&	0.00	&	0.10	&	0.14	&	0.71	&	0.87	&	1.00	&	0.03	&	0.07	&	0.69	&	0.87	&	1.00	&	0.03	&	0.07	&	0.78	&	0.91	&	1.00	&	0.03	&	0.07	\\
\rowcolor{gray!10}
	&	$	VaR \quad (\alpha =  0.025)	$	&	0.01	&	0.01	&	0.03	&	0.03	&	0.08	&	0.85	&	0.19	&	0.52	&	0.02	&	0.07	&	0.77	&	0.17	&	0.47	&	0.02	&	0.07	&	0.64	&	0.26	&	0.59	&	0.03	&	0.07	\\
\rowcolor{gray!10}
			&	$	CoVaR \quad (\beta =  0.05)	$	&	0.00	&	0.00	&	0.00	&	0.16	&	0.19	&	0.54	&	0.74	&	1.00	&	0.04	&	0.07	&	0.56	&	0.75	&	1.00	&	0.04	&	0.07	&	0.25	&	0.28	&	0.34	&	0.08	&	0.11	\\
\rowcolor{gray!10}
\multirow{-3}{*}{F:BNP}			&	$	CoVaR \quad (\beta =  0.025)	$	&	0.00	&	0.00	&	0.00	&	0.11	&	0.17	&	0.96	&	0.95	&	1.00	&	0.02	&	0.07	&	0.97	&	0.95	&	1.00	&	0.02	&	0.06	&	0.88	&	0.94	&	1.00	&	0.02	&	0.06	\\
\multirow{3}{*}{F:SGE}	&	$	VaR \quad (\alpha =  0.025)	$	&	0.22	&	0.21	&	0.21	&	0.03	&	0.08	&	0.60	&	0.34	&	0.60	&	0.02	&	0.08	&	0.53	&	0.30	&	0.55	&	0.02	&	0.08	&	0.97	&	0.50	&	0.51	&	0.03	&	0.08	\\
			&	$	CoVaR \quad (\beta =  0.05)	$	&	0.00	&	0.00	&	0.00	&	0.17	&	0.20	&	0.59	&	0.77	&	1.00	&	0.04	&	0.07	&	0.98	&	0.81	&	0.97	&	0.05	&	0.08	&	0.41	&	0.45	&	0.41	&	0.07	&	0.10	\\
			&	$	CoVaR \quad (\beta =  0.025)	$	&	0.00	&	0.00	&	0.00	&	0.10	&	0.15	&	1.00	&	0.95	&	1.00	&	0.03	&	0.07	&	0.99	&	0.95	&	1.00	&	0.03	&	0.07	&	0.93	&	0.95	&	1.00	&	0.02	&	0.06	\\
\rowcolor{gray!10}
	&	$	VaR \quad (\alpha =  0.025)	$	&	0.37	&	0.46	&	0.45	&	0.03	&	0.08	&	0.89	&	0.52	&	0.27	&	0.03	&	0.08	&	0.94	&	0.48	&	0.44	&	0.02	&	0.08	&	0.43	&	0.69	&	0.58	&	0.03	&	0.08	\\
\rowcolor{gray!10}
			&	$	CoVaR \quad (\beta =  0.05)	$	&	0.00	&	0.00	&	0.00	&	0.16	&	0.19	&	0.74	&	0.69	&	0.99	&	0.06	&	0.09	&	0.70	&	0.67	&	0.99	&	0.06	&	0.09	&	0.28	&	0.31	&	0.60	&	0.08	&	0.10	\\
\rowcolor{gray!10}
\multirow{-3}{*}{F:CRDA}			&	$	CoVaR \quad (\beta =  0.025)	$	&	0.00	&	0.00	&	0.00	&	0.12	&	0.20	&	0.58	&	0.77	&	0.98	&	0.03	&	0.07	&	0.55	&	0.75	&	0.98	&	0.04	&	0.07	&	0.65	&	0.82	&	0.99	&	0.03	&	0.07	\\
\multirow{3}{*}{E:BBVA}	&	$	VaR \quad (\alpha =  0.025)	$	&	0.00	&	0.00	&	0.00	&	0.04	&	0.08	&	0.31	&	0.56	&	0.19	&	0.03	&	0.07	&	0.31	&	0.56	&	0.61	&	0.03	&	0.07	&	0.02	&	0.06	&	0.13	&	0.03	&	0.07	\\
			&	$	CoVaR \quad (\beta =  0.05)	$	&	0.01	&	0.04	&	0.07	&	0.11	&	0.13	&	0.73	&	0.79	&	0.99	&	0.04	&	0.07	&	0.73	&	0.79	&	0.99	&	0.04	&	0.07	&	0.76	&	0.66	&	0.96	&	0.06	&	0.09	\\
			&	$	CoVaR \quad (\beta =  0.025)	$	&	0.00	&	0.01	&	0.01	&	0.07	&	0.10	&	0.68	&	0.83	&	1.00	&	0.03	&	0.07	&	0.68	&	0.83	&	1.00	&	0.03	&	0.07	&	0.83	&	0.89	&	1.00	&	0.03	&	0.06	\\
\rowcolor{gray!10}
	&	$	VaR \quad (\alpha =  0.025)	$	&	0.00	&	0.01	&	0.00	&	0.03	&	0.07	&	0.27	&	0.26	&	0.14	&	0.03	&	0.07	&	0.31	&	0.31	&	0.16	&	0.03	&	0.07	&	0.06	&	0.07	&	0.12	&	0.03	&	0.07	\\
\rowcolor{gray!10}
			&	$	CoVaR \quad (\beta =  0.05)	$	&	0.00	&	0.00	&	0.00	&	0.14	&	0.16	&	0.72	&	0.78	&	0.99	&	0.04	&	0.07	&	0.73	&	0.79	&	0.99	&	0.04	&	0.07	&	0.41	&	0.42	&	0.77	&	0.07	&	0.09	\\
\rowcolor{gray!10}
\multirow{-3}{*}{E:SAN}			&	$	CoVaR \quad (\beta =  0.025)	$	&	0.00	&	0.00	&	0.00	&	0.10	&	0.13	&	0.69	&	0.84	&	1.00	&	0.03	&	0.07	&	0.68	&	0.83	&	1.00	&	0.03	&	0.06	&	0.78	&	0.88	&	1.00	&	0.03	&	0.06	\\
\multirow{3}{*}{E:BSAB}	&	$	VaR \quad (\alpha =  0.025)	$	&	0.00	&	0.00	&	0.00	&	0.03	&	0.08	&	0.00	&	0.01	&	0.00	&	0.03	&	0.07	&	0.01	&	0.02	&	0.00	&	0.03	&	0.07	&	0.05	&	0.09	&	0.00	&	0.03	&	0.07	\\
			&	$	CoVaR \quad (\beta =  0.05)	$	&	0.11	&	0.26	&	0.04	&	0.09	&	0.11	&	0.45	&	0.68	&	0.98	&	0.04	&	0.07	&	0.48	&	0.69	&	0.99	&	0.04	&	0.07	&	0.59	&	0.77	&	0.99	&	0.04	&	0.07	\\
			&	$	CoVaR \quad (\beta =  0.025)	$	&	0.04	&	0.07	&	0.00	&	0.06	&	0.09	&	0.92	&	0.94	&	1.00	&	0.03	&	0.07	&	0.89	&	0.94	&	1.00	&	0.03	&	0.06	&	0.40	&	0.63	&	1.00	&	0.04	&	0.08	\\
\rowcolor{gray!10}
	&	$	VaR \quad (\alpha =  0.025)	$	&	0.02	&	0.01	&	0.00	&	0.03	&	0.09	&	0.19	&	0.09	&	0.00	&	0.03	&	0.09	&	0.15	&	0.09	&	0.00	&	0.03	&	0.09	&	0.31	&	0.43	&	0.04	&	0.03	&	0.08	\\
\rowcolor{gray!10}
			&	$	CoVaR \quad (\beta =  0.05)	$	&	0.00	&	0.00	&	0.00	&	0.14	&	0.17	&	0.68	&	0.77	&	0.99	&	0.04	&	0.08	&	0.67	&	0.77	&	0.99	&	0.04	&	0.07	&	0.89	&	0.75	&	0.94	&	0.05	&	0.08	\\
\rowcolor{gray!10}
\multirow{-3}{*}{B:KB}			&	$	CoVaR \quad (\beta =  0.025)	$	&	0.00	&	0.00	&	0.00	&	0.09	&	0.13	&	0.78	&	0.92	&	1.00	&	0.02	&	0.06	&	0.76	&	0.92	&	1.00	&	0.02	&	0.06	&	0.81	&	0.93	&	1.00	&	0.02	&	0.06	\\
\multirow{3}{*}{H:INGA}	&	$	VaR \quad (\alpha =  0.025)	$	&	0.15	&	0.29	&	0.28	&	0.03	&	0.08	&	0.68	&	0.38	&	0.64	&	0.02	&	0.08	&	0.68	&	0.38	&	0.64	&	0.02	&	0.08	&	0.89	&	0.24	&	0.27	&	0.03	&	0.08	\\
			&	$	CoVaR \quad (\beta =  0.05)	$	&	0.00	&	0.00	&	0.00	&	0.14	&	0.17	&	0.25	&	0.49	&	0.99	&	0.02	&	0.06	&	0.25	&	0.49	&	0.99	&	0.02	&	0.05	&	0.74	&	0.69	&	0.91	&	0.06	&	0.09	\\
			&	$	CoVaR \quad (\beta =  0.025)	$	&	0.00	&	0.00	&	0.00	&	0.10	&	0.15	&	0.99	&	0.95	&	1.00	&	0.02	&	0.06	&	0.99	&	0.95	&	1.00	&	0.02	&	0.06	&	0.92	&	0.95	&	1.00	&	0.02	&	0.06	\\

\bottomrule
			
\end{tabular}
\caption[]{\label{tab:0025}\footnotesize $P$-values of the LR test of unconditional coverage (uc), coverage and independence (cc) and dynamic quantile tests (dq) for both $VaR$ and $CoVaR$  with $\alpha = 0.025$ for all analyzed models and banks. The values of the magnitude (LM) and of the asymmetric loss (LA) functions are also reported. }
\end{scriptsize}
\end{center}
\end{sidewaystable}


\begin{sidewaystable}
\begin{center}
\begin{scriptsize}
\begin{tabular}{ll|ccccc|ccccc|ccccc|ccccc}
\toprule

			&				&	\multicolumn{5}{c}{MNormal}									&	\multicolumn{5}{c}{\textbf{MGH}}									&	\multicolumn{5}{c}{MNTS}									&	\multicolumn{5}{c}{Copula}									\\
			
&				&	uc	&	cc	&	dq	&	LM	&	LA	&	uc	&	cc	&	dq	&	LM	&	LA	&	uc	&	cc	&	dq	&	LM	&	LA	&	uc	&	cc	&	dq	&	LM	&	LA	\\

\midrule
\rowcolor{gray!10}
&	$	VaR \quad (\alpha =  0.01)	$	&	0.00	&	0.00	&	0.00	&	0.02	&	0.07	&	0.23	&	0.40	&	0.66	&	0.01	&	0.07	&	0.23	&	0.40	&	0.66	&	0.01	&	0.07	&	0.49	&	0.51	&	0.61	&	0.01	&	0.07	\\
\rowcolor{gray!10}
\multirow{-2}{*}{D:DBK}				&	$	CoVaR \quad (\beta =  0.05)	$	&	0.00	&	0.00	&	0.00	&	0.16	&	0.20	&	0.52	&	0.69	&	0.52	&	0.07	&	0.12	&	0.52	&	0.69	&	0.52	&	0.07	&	0.11	&	0.45	&	0.63	&	1.00	&	0.08	&	0.11	\\
\multirow{2}{*}{D:CBK}	&	$	VaR \quad (\alpha =  0.01)	$	&	0.00	&	0.00	&	0.00	&	0.02	&	0.08	&	0.13	&	0.18	&	0.57	&	0.01	&	0.08	&	0.13	&	0.18	&	0.56	&	0.01	&	0.08	&	0.39	&	0.44	&	0.67	&	0.01	&	0.08	\\
			&	$	CoVaR \quad (\beta =  0.05)	$	&	0.00	&	0.00	&	0.00	&	0.19	&	0.26	&	0.57	&	0.73	&	0.58	&	0.07	&	0.10	&	0.57	&	0.73	&	0.58	&	0.07	&	0.10	&	0.18	&	0.29	&	0.01	&	0.10	&	0.14	\\
\rowcolor{gray!10}
&	$	VaR \quad (\alpha =  0.01)	$	&	0.00	&	0.00	&	0.00	&	0.02	&	0.08	&	0.11	&	0.23	&	0.43	&	0.01	&	0.08	&	0.16	&	0.30	&	0.44	&	0.01	&	0.08	&	0.49	&	0.61	&	0.58	&	0.01	&	0.08	\\
\rowcolor{gray!10}
\multirow{-2}{*}{I:UCG}				&	$	CoVaR \quad (\beta =  0.05)	$	&	0.01	&	0.01	&	0.00	&	0.15	&	0.18	&	0.53	&	0.68	&	0.98	&	0.08	&	0.12	&	0.56	&	0.71	&	0.98	&	0.08	&	0.11	&	0.69	&	0.80	&	0.99	&	0.07	&	0.10	\\
\multirow{2}{*}{I:ISP}	&	$	VaR \quad (\alpha =  0.01)	$	&	0.00	&	0.00	&	0.00	&	0.02	&	0.08	&	0.30	&	0.37	&	0.01	&	0.01	&	0.08	&	0.30	&	0.37	&	0.01	&	0.01	&	0.08	&	0.07	&	0.10	&	0.00	&	0.01	&	0.08	\\
			&	$	CoVaR \quad (\beta =  0.05)	$	&	0.00	&	0.00	&	0.00	&	0.17	&	0.21	&	1.00	&	0.90	&	0.99	&	0.05	&	0.08	&	1.00	&	0.90	&	0.99	&	0.05	&	0.08	&	0.62	&	0.77	&	0.53	&	0.07	&	0.10	\\
\rowcolor{gray!10}
	&	$	VaR \quad (\alpha =  0.01)	$	&	0.00	&	0.00	&	0.00	&	0.02	&	0.07	&	0.74	&	0.11	&	0.00	&	0.01	&	0.07	&	0.74	&	0.11	&	0.00	&	0.01	&	0.07	&	0.60	&	0.16	&	0.01	&	0.01	&	0.07	\\
\rowcolor{gray!10}
\multirow{-2}{*}{F:BNP}			&	$	CoVaR \quad (\beta =  0.05)	$	&	0.00	&	0.01	&	0.00	&	0.16	&	0.19	&	0.75	&	0.83	&	0.97	&	0.06	&	0.10	&	0.75	&	0.83	&	0.97	&	0.06	&	0.10	&	0.91	&	0.88	&	1.00	&	0.05	&	0.09	\\
\multirow{2}{*}{F:SGE}	&	$	VaR \quad (\alpha =  0.01)	$	&	0.01	&	0.02	&	0.02	&	0.01	&	0.08	&	0.39	&	0.54	&	0.85	&	0.01	&	0.08	&	0.49	&	0.61	&	0.84	&	0.01	&	0.08	&	0.74	&	0.70	&	0.82	&	0.01	&	0.08	\\
			&	$	CoVaR \quad (\beta =  0.05)	$	&	0.00	&	0.01	&	0.00	&	0.16	&	0.20	&	0.69	&	0.89	&	1.00	&	0.03	&	0.08	&	0.66	&	0.87	&	1.00	&	0.03	&	0.07	&	0.60	&	0.84	&	1.00	&	0.03	&	0.07	\\
\rowcolor{gray!10}
	&	$	VaR \quad (\alpha =  0.01)	$	&	0.00	&	0.00	&	0.00	&	0.02	&	0.07	&	0.85	&	0.66	&	0.08	&	0.01	&	0.07	&	0.85	&	0.66	&	0.08	&	0.01	&	0.07	&	0.60	&	0.63	&	0.68	&	0.01	&	0.07	\\
\rowcolor{gray!10}
\multirow{-2}{*}{F:CRDA}			&	$	CoVaR \quad (\beta =  0.05)	$	&	0.00	&	0.00	&	0.00	&	0.20	&	0.26	&	0.38	&	0.56	&	0.45	&	0.09	&	0.12	&	0.38	&	0.56	&	0.45	&	0.09	&	0.12	&	0.42	&	0.61	&	0.90	&	0.08	&	0.11	\\
\multirow{2}{*}{E:BBVA}	&	$	VaR \quad (\alpha =  0.01)	$	&	0.00	&	0.00	&	0.00	&	0.02	&	0.07	&	0.98	&	0.71	&	0.86	&	0.01	&	0.06	&	0.85	&	0.68	&	0.84	&	0.01	&	0.06	&	0.18	&	0.24	&	0.56	&	0.01	&	0.06	\\
			&	$	CoVaR \quad (\beta =  0.05)	$	&	0.02	&	0.02	&	0.00	&	0.13	&	0.15	&	0.35	&	0.53	&	0.99	&	0.09	&	0.12	&	0.38	&	0.56	&	0.99	&	0.09	&	0.12	&	0.55	&	0.66	&	0.53	&	0.07	&	0.10	\\
\rowcolor{gray!10}
	&	$	VaR \quad (\alpha =  0.01)	$	&	0.00	&	0.00	&	0.00	&	0.02	&	0.07	&	0.39	&	0.54	&	0.04	&	0.01	&	0.06	&	0.39	&	0.54	&	0.04	&	0.01	&	0.06	&	0.49	&	0.51	&	0.15	&	0.01	&	0.07	\\
\rowcolor{gray!10}
\multirow{-2}{*}{E:SAN}			&	$	CoVaR \quad (\beta =  0.05)	$	&	0.00	&	0.00	&	0.00	&	0.17	&	0.21	&	0.25	&	0.40	&	0.51	&	0.10	&	0.14	&	0.25	&	0.40	&	0.52	&	0.10	&	0.14	&	0.45	&	0.63	&	0.53	&	0.08	&	0.11	\\
\multirow{2}{*}{E:BSAB}	&	$	VaR \quad (\alpha =  0.01)	$	&	0.00	&	0.00	&	0.00	&	0.02	&	0.07	&	0.00	&	0.01	&	0.00	&	0.02	&	0.07	&	0.00	&	0.01	&	0.00	&	0.02	&	0.07	&	0.72	&	0.64	&	0.04	&	0.01	&	0.07	\\
			&	$	CoVaR \quad (\beta =  0.05)	$	&	0.10	&	0.21	&	0.00	&	0.10	&	0.12	&	0.65	&	0.87	&	0.99	&	0.04	&	0.08	&	0.65	&	0.87	&	0.99	&	0.04	&	0.07	&	0.40	&	0.58	&	0.10	&	0.08	&	0.12	\\
\rowcolor{gray!10}
	&	$	VaR \quad (\alpha =  0.01)	$	&	0.00	&	0.00	&	0.00	&	0.02	&	0.08	&	0.23	&	0.40	&	0.19	&	0.01	&	0.08	&	0.23	&	0.40	&	0.19	&	0.01	&	0.08	&	0.98	&	0.65	&	0.71	&	0.01	&	0.08	\\
\rowcolor{gray!10}
\multirow{-2}{*}{B:KB}			&	$	CoVaR \quad (\beta =  0.05)	$	&	0.03	&	0.04	&	0.00	&	0.13	&	0.16	&	0.97	&	0.90	&	0.98	&	0.05	&	0.09	&	0.97	&	0.90	&	0.98	&	0.05	&	0.09	&	0.55	&	0.81	&	1.00	&	0.03	&	0.07	\\
\multirow{2}{*}{H:INGA}	&	$	VaR \quad (\alpha =  0.01)	$	&	0.00	&	0.00	&	0.00	&	0.02	&	0.07	&	0.49	&	0.43	&	0.49	&	0.01	&	0.08	&	0.49	&	0.43	&	0.49	&	0.01	&	0.08	&	0.72	&	0.66	&	0.68	&	0.01	&	0.07	\\
			&	$	CoVaR \quad (\beta =  0.05)	$	&	0.01	&	0.01	&	0.00	&	0.15	&	0.17	&	0.69	&	0.80	&	0.99	&	0.07	&	0.10	&	0.69	&	0.80	&	1.00	&	0.07	&	0.10	&	0.88	&	0.88	&	1.00	&	0.06	&	0.09	\\

\bottomrule
			
\end{tabular}
\caption[]{\label{tab:001}\footnotesize $P$-values of the LR test of unconditional coverage (uc), coverage and independence (cc) and dynamic quantile tests (dq) for both $VaR$ and $CoVaR$  with $\alpha = 0.01$ for all analyzed models and banks. The values of the magnitude (LM) and of the asymmetric loss (LA) functions are also reported. }
\end{scriptsize}
\end{center}
\end{sidewaystable}


\begin{sidewaystable}
\begin{center}
\begin{footnotesize}
\begin{tabular}{l|K{1.4cm}K{1.4cm}K{1.4cm}K{1.4cm}K{1.4cm}|K{1.4cm}K{1.4cm}K{1.4cm}K{1.4cm}K{1.4cm}K{1.4cm}|K{1.4cm}}
\toprule
\multirow{2}{*}{ticker}   &\multicolumn{5}{c}{$\Delta CoVaR$ indicators} &\multicolumn{6}{c}{GSIBs indicators and scores} & \\
 & Normal & MNormal & MGH & MNTS & Copula & Size & Inte & Subs & Comp & Cja & Score & Ad.score\\
\midrule
D:DBK & \cellcolor[HTML]{E3FE50} 946 & \cellcolor[HTML]{E5FE4C} 968 & \cellcolor[HTML]{EFFE3B} 991 & \cellcolor[HTML]{EFFE3C} 990 & \cellcolor[HTML]{EBFE43} 987 & \cellcolor[HTML]{FFBF00}1257 & \cellcolor[HTML]{FF9F00}1373 & \cellcolor[HTML]{FF4500}2902 & \cellcolor[HTML]{FF4500}3149 & \cellcolor[HTML]{FFA900}1398 & \cellcolor[HTML]{FF4500}2016 & \cellcolor[HTML]{FFA000}1352\\
F:BNP & \cellcolor[HTML]{FFD300}1085 & \cellcolor[HTML]{FFE000}1048 & \cellcolor[HTML]{FFE400}1030 & \cellcolor[HTML]{FFE400}1029 & \cellcolor[HTML]{FFC500}1029 & \cellcolor[HTML]{FF4500}1662 & \cellcolor[HTML]{FF8500}1450 & \cellcolor[HTML]{FF8A00}2156 & \cellcolor[HTML]{FF9100}2237 & \cellcolor[HTML]{FF4500}1855 & \cellcolor[HTML]{FF6500}1872 & \cellcolor[HTML]{FF6F00}1488\\
F:SGE & \cellcolor[HTML]{FFF100}1043 & \cellcolor[HTML]{FFF200}1030 & \cellcolor[HTML]{FFF500}1018 & \cellcolor[HTML]{FFF400}1018 & \cellcolor[HTML]{FFE100}1017 & \cellcolor[HTML]{FFEE00}1072 & \cellcolor[HTML]{FFE900}1127 & \cellcolor[HTML]{FFC400}1362 & \cellcolor[HTML]{FFCB00}1359 & \cellcolor[HTML]{FCFF17} 869 & \cellcolor[HTML]{FFD600}1158 & \cellcolor[HTML]{FFEB00}1117\\
F:CRDA & \cellcolor[HTML]{E9FE47} 962 & \cellcolor[HTML]{EEFE3D} 984 & \cellcolor[HTML]{ECFE41} 987 & \cellcolor[HTML]{EDFE3F} 988 & \cellcolor[HTML]{EDFE3F} 989 & \cellcolor[HTML]{FFAA00}1339 & \cellcolor[HTML]{FFD800}1187 & \cellcolor[HTML]{FFD500}1108 & \cellcolor[HTML]{FFEC00} 840 & \cellcolor[HTML]{EAFE45} 669 & \cellcolor[HTML]{FFE900}1029 & \cellcolor[HTML]{FFCC00}1214\\
E:SAN & \cellcolor[HTML]{FFCE00}1091 & \cellcolor[HTML]{FFDB00}1053 & \cellcolor[HTML]{FFE100}1032 & \cellcolor[HTML]{FFE100}1031 & \cellcolor[HTML]{FFC500}1029 & \cellcolor[HTML]{FFC600}1230 & \cellcolor[HTML]{FFF700}1081 & \cellcolor[HTML]{FFFF00} 492 & \cellcolor[HTML]{FBFF1B} 402 & \cellcolor[HTML]{FF6700}1725 & \cellcolor[HTML]{FFEF00} 986 & \cellcolor[HTML]{FFED00}1108\\
I:UCG & \cellcolor[HTML]{EAFE44} 966 & \cellcolor[HTML]{F0FE3A} 987 & \cellcolor[HTML]{ECFE41} 987 & \cellcolor[HTML]{EDFE3F} 988 & \cellcolor[HTML]{DAFE5C} 974 & \cellcolor[HTML]{EBFE43} 845 & \cellcolor[HTML]{FBFF1B}1025 & \cellcolor[HTML]{FEFF0A} 459 & \cellcolor[HTML]{FFFF04} 521 & \cellcolor[HTML]{FFF900} 944 & \cellcolor[HTML]{F7FF29} 759 & \cellcolor[HTML]{F6FF2B} 993\\
H:INGA & \cellcolor[HTML]{FFF100}1043 & \cellcolor[HTML]{FFF400}1028 & \cellcolor[HTML]{FFEF00}1022 & \cellcolor[HTML]{FFEE00}1022 & \cellcolor[HTML]{FFD100}1024 & \cellcolor[HTML]{F6FF2B} 931 & \cellcolor[HTML]{DFFE55} 850 & \cellcolor[HTML]{FDFF13} 418 & \cellcolor[HTML]{F9FF23} 341 & \cellcolor[HTML]{FFCC00}1202 & \cellcolor[HTML]{F6FF2B} 748 & \cellcolor[HTML]{E1FE53} 865\\
D:CBK & \cellcolor[HTML]{98FB98} 792 & \cellcolor[HTML]{98FB98} 854 & \cellcolor[HTML]{98FB98} 902 & \cellcolor[HTML]{98FB98} 904 & \cellcolor[HTML]{98FB98} 933 & \cellcolor[HTML]{B2FC85} 472 & \cellcolor[HTML]{C2FD77} 685 & \cellcolor[HTML]{FFFF00} 492 & \cellcolor[HTML]{FFFE00} 552 & \cellcolor[HTML]{CAFD6E} 354 & \cellcolor[HTML]{E4FE4E} 511 & \cellcolor[HTML]{B7FC81} 646\\
E:BBVA & \cellcolor[HTML]{FFDD00}1071 & \cellcolor[HTML]{FFE600}1042 & \cellcolor[HTML]{FFE900}1026 & \cellcolor[HTML]{FFEA00}1025 & \cellcolor[HTML]{FFCC00}1026 & \cellcolor[HTML]{CCFD6D} 626 & \cellcolor[HTML]{98FB98} 498 & \cellcolor[HTML]{F9FF21} 323 & \cellcolor[HTML]{F6FF2C} 239 & \cellcolor[HTML]{EDFE3E} 708 & \cellcolor[HTML]{E1FE52} 479 & \cellcolor[HTML]{98FB98} 521\\
I:ISP & \cellcolor[HTML]{F8FF26}1002 & \cellcolor[HTML]{F9FF20}1006 & \cellcolor[HTML]{FAFF1E}1005 & \cellcolor[HTML]{FAFF20}1004 & \cellcolor[HTML]{F0FE3A} 991 & \cellcolor[HTML]{C2FD76} 566 & \cellcolor[HTML]{C9FD6F} 724 & \cellcolor[HTML]{F8FF25} 287 & \cellcolor[HTML]{F9FF20} 360 & \cellcolor[HTML]{C2FD77} 277 & \cellcolor[HTML]{DFFE56} 443 & \cellcolor[HTML]{C1FD78} 695\\
\bottomrule
			
\end{tabular}
\caption[]{\label{tab:GSIBs}\footnotesize Average indicators and scores over the years from 2013 to 2018. The adjusted score is computed on the basis of the weights minimizing the average relative percentage error between the GSIBs and the $\Delta CoVaR$ score in the MNTS case. For the $\Delta CoVaR$, we consider $\alpha$ and $\beta$ equal to 0.05.}
\end{footnotesize}
\end{center}
\end{sidewaystable}

\end{document}